\documentclass[11pt,twoside,a4paper,cmspaper,final,collab]{cms-tdr}
\def\svnVersion{a45a732-D}\def\svnDate{2026/07/28}\def\cmsCernNoTag{CERN-EP-2026-183}\def\cmsCernDate{\today}\def\cmsMessage{Submitted to the Journal of High Energy Physics}
\begin{document}\cmsNoteHeader{SMP-25-005}

\newcommand{\PGcbDiP}[1]{{\HepParticleResonanceFull{\PGc}{b\mathit{i}}{}{#1}{}{}}\xspace}
\newcommand{\pp}{\ensuremath{\Pp\Pp}\xspace}
\newcommand{\ptrel}{\ensuremath{\pt^{\text{rel}}}\xspace}
\newcommand{\ptjet}{\ensuremath{\pt^{\text{jet}}}\xspace}
\newcommand{\vecpjet}{\ensuremath{\vec{p}_\text{jet}}\xspace}
\newcommand{\vecpups}{\ensuremath{\vec{p}_{\PGU}}\xspace}

\cmsNoteHeader{SMP-25-005}
\title{Measurement of the fragmentation properties of jets containing \texorpdfstring{\PGUPnS}{Upsilon(nS)} mesons in proton-proton collisions at \texorpdfstring{$\sqrt{s}=13\TeV$}{sqrt(s) = 13 TeV}}
\date{\today}

\abstract{A measurement of the fragmentation properties of jets containing \PGUPnS mesons using proton-proton collision data at $\sqrt{s}=13\TeV$, corresponding to an integrated luminosity of 138\fbinv, is presented. The \PGUPnS mesons associated with jets are reconstructed via their decays producing pairs of oppositely charged muons. The longitudinal and transverse projections of the momenta of the \PGUPnS mesons along the momenta of the corresponding jets are studied. The results are compared with Monte Carlo predictions, including recent developments in the modelling of quarkonia production in parton showers. The description of the data by these Monte Carlo predictions is found to be unsatisfactory, leaving room for improvements in the modelling of such processes.}

\hypersetup{%
pdfauthor={CMS Collaboration},%
pdftitle={Measurement of the fragmentation properties of jets containing Upsilon mesons in proton-proton collisions at sqrt(s) = 13 TeV},%
pdfsubject={CMS},%
pdfkeywords={CMS, QCD, NRQCD, jets, Upsilon, quarkonia}}

\maketitle

\section{Introduction}

\begin{figure}[!h]
\centering
\includegraphics[width=0.3\textwidth]{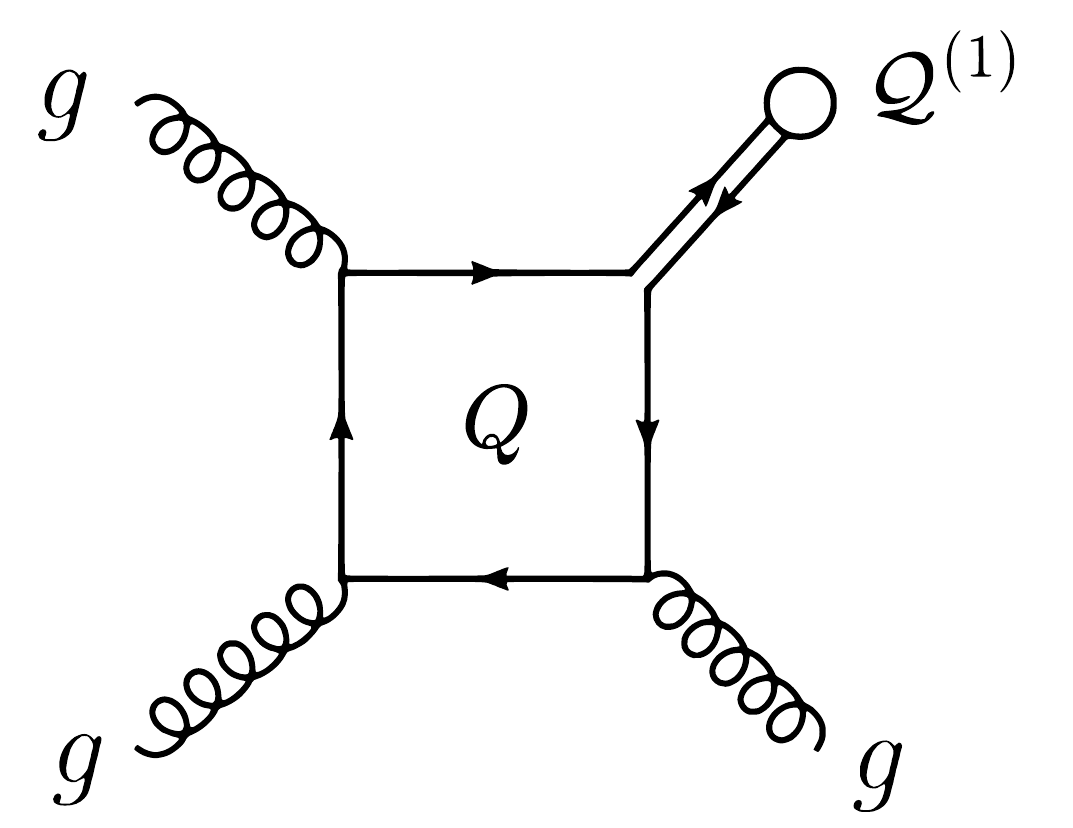}
\hspace{0.2\textwidth}
\includegraphics[width=0.3\textwidth]{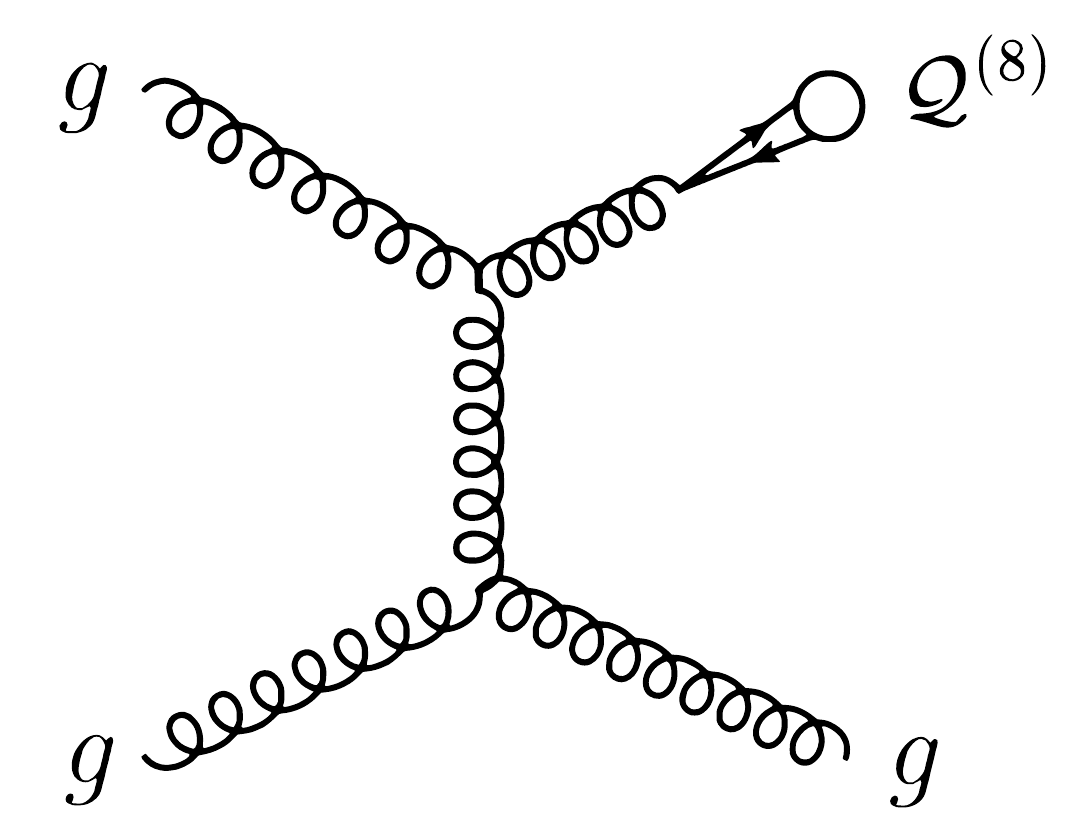}
\caption{Example diagrams for the hard production of \PGUPnS mesons in the colour singlet (left) and colour octet mechanisms (right).}
\label{fig:diagrams}
\end{figure}

The production of quarkonia, bound states of charm and bottom quarks with their respective anti-quarks, involves both perturbative and non-perturbative aspects of quantum chromodynamics (QCD). The large mass of the heavy quark provides a hard scale that allows the production of the initial heavy quark-antiquark pair to be described perturbatively. The subsequent evolution into a physical quarkonium state involves long-distance, non-perturbative dynamics. A widely used theoretical framework to describe this process is non-relativistic QCD (NRQCD)~\cite{caswell, bodwin, Cho96_1, Cho96_2}, an effective field theory in which the production cross section is factorized into short-distance and long-distance contributions. The short-distance contributions are calculable within perturbative QCD, whereas the long-distance contributions are parameterised in terms of so-called long-distance matrix elements (LDMEs), which encode the non-perturbative transition of the heavy-quark pair into the observed quarkonium state. The differential cross section for the production of a quarkonium state $\mathcal{Q}$ can thus be expressed as
\begin{equation}\label{eq:ldmeSigma}
    \rd\sigma[\pp\to\mathcal{Q}\PX] = \sum_i\langle\mathcal{O}_i[\mathcal{Q}]\rangle \rd\sigma[\pp\to\PQQ\PAQQ(i)\PX],
\end{equation}
where \PQQ is a heavy quark, charm or bottom, and $\rd\sigma[\pp\to\PQQ\PAQQ(i)\PX]$ is the differential cross section for the production of a Fock state $i\equiv{}^{2S+1}L_J^{(c)}$ with well-defined orbital angular momentum $L$, spin $S$ and total angular momentum $J$. The index $c$ denotes the colour quantum number and takes the value of 1 for colour-singlet and 8 for colour-octet states, whose production is illustrated in Fig.~\ref{fig:diagrams}. The coefficients $\langle\mathcal{O}_i[\mathcal{Q}]\rangle$ are the LDMEs described above~\cite{bodwin}.

Early measurements of charmonium and bottomonium cross sections at the Tevatron~\cite{CDF:1997ykw, CDF:1995gwi, D0:1996awi, D0:2005klj} revealed production rates larger than those predicted by the colour-singlet mechanism alone, motivating the introduction of the colour octet mechanism. Later measurements at the CERN LHC~\cite{CMS:2017dju, ATLAS:2023qnh, LHCb:2015foc, ATLAS:2012lmu, CMS:2026ccg, LHCb:2018yzj} extended these results to higher transverse momenta and confirmed the success of the NRQCD factorization framework to describe the quarkonia production cross sections.

Additional stringent tests of NRQCD are provided by quarkonium polarization measurements. In NRQCD, quarkonia states produced at high transverse momentum are expected to exhibit predominantly transverse polarization~\cite{Braaten:1999qk, Beneke:1996yw}. However, measurements of \PJGy and \PGUPnS polarization at the Tevatron~\cite{CDF:2007msx, D0:2008yos, CDF:2011ag} and the LHC~\cite{CMS:2013gbz, CMS:2024igk, cmsUpsiPol, lhcbJpsiPol, LHCb:2017scf} show very mild transverse polarizations over the full range of transverse momenta. This longstanding tension between theoretical predictions and experimental observations, known as the quarkonia polarization puzzle, remains one of the central open questions for the understanding of quarkonium production.

The production of heavy-flavoured hadrons in jets provides a useful probe of QCD in a regime which cannot be described perturbatively. When the hadron momentum can be fully reconstructed from its charged decay products, its properties can be studied through the momentum it carries relative to the jet containing it. These observables are sensitive to the fragmentation functions $D_{\PQq}^{\Ph}(z,Q^2)$~\cite{mele1991, cacciari2003}, which describe the probability for a hadron \Ph to carry a fraction $z$ of the energy of the parent quark \PQq at a scale $Q^2$.

Despite their simplicity, only a few measurements of heavy-quark fragmentation have been produced so far at the LHC~\cite{atlasDpm, atlasBpm, aliceD0, LHCb:2026mbl}. Among these measurements, the fragmentation properties of jets containing quarkonia are of particular importance, as they provide an alternative methodology to study the polarization properties of these objects~\cite{Baumgart:2014upa, zKang}. Recent measurements of the fragmentation of jets containing quarkonia at the LHC~\cite{lhcbJpsiFrag,lhcbPsi2SFrag,cmsJpsiJet,cmsJpsiFrag} show that the predictions by \PYTHIA8.2~\cite{pythia82} and \PYTHIA8.3~\cite{pythia83} underestimate the amount of additional hadronic activity produced inside the jets containing \PJGy mesons. These discrepancies have been shown to have an important dependence on the long-distance aspects of the NRQCD modelling of charmonium production~\cite{Bain:2017wvk, Wang:2025drz}. These observations have triggered theoretical developments on quarkonia production from parton showers to try to address the disagreement~\cite{cooke23}, as well as explicit calculations of quarkonia fragmentation functions based on the collinear factorization approach for vector and pseudoscalar states~\cite{celiberto24, celiberto25}.

This work presents measurements of the fragmentation of jets containing \PGUPnS mesons using data collected in 2016--2018 by the CMS experiment in proton-proton (\pp) collisions at $\sqrt{s}=13\TeV$, corresponding to an integrated luminosity of 138\fbinv. These measurements extend the above-mentioned results to the bottomonium sector by considering the \PGUP{1S}, \PGUP{2S}, and \PGUP{3S} states in different kinematic ranges, as well as including the transverse profiles of the quarkonium momentum, enhancing sensitivity to transverse-momentum-dependent distributions~\cite{scimemi1, scimemi2, copeland, Chapon:2020heu}. More explicitly, observables considered are
\begin{equation}\label{eq:definition}
    z = \frac{\vecpjet\cdot\vecpups}{\big\vert\vecpjet\big\vert^2};
    \qquad
    \ptrel = \frac{\big\vert\vecpjet\times\vecpups\big\vert}{\big\vert\vecpjet\big\vert}.
\end{equation}
In Eq.~\eqref{eq:definition}, the value of \vecpups is the three-momentum of the \PGU meson, whereas \vecpjet is the three-momentum of the jet containing it. Thus, $z$ is a measure of the longitudinal component of the momentum fraction carried by the \PGU along the direction of the jet axis, while \ptrel quantifies the momentum in the direction orthogonal to the jet axis.

The present paper is structured as follows: the CMS detector is described in Section~\ref{sec:detector}, followed by a description of the Monte Carlo (MC) predictions used for the analysis in Section~\ref{sec:montecarlo}. Section~\ref{sec:data} gives a summary of the data and the event selection criteria used in the analysis. The description of the fits used to determine the yields of the \PGUPnS samples is given in Section~\ref{sec:purity}, and the corrections for detector effects are described in Section~\ref{sec:unfolding}. Section~\ref{sec:systematics} is dedicated to the estimation of the systematic uncertainties and Section~\ref{sec:results} contains an overview of the experimental results and their comparison to particle-level MC predictions. Finally, Section~\ref{sec:summary} summarizes the results. Tabulated results can be found in the HEPData record for this analysis~\cite{hepdata}.

\section{The CMS detector and object reconstruction}
\label{sec:detector}

The CMS apparatus~\cite{CMS:2008xjf,CMS:2023gfb} is a multipurpose, nearly hermetic detector, designed to trigger on~\cite{CMS:2020cmk,CMS:2016ngn,CMS:2024aqx} and identify electrons, muons, photons, and (charged and neutral) hadrons~\cite{CMS:2020uim,CMS:2018rym,CMS:2014pgm}. Its central feature is a superconducting solenoid of 6\unit{m} internal diameter, providing a magnetic field of 3.8\unit{T}. Within the solenoid volume are a silicon pixel and strip tracker, a lead tungstate crystal electromagnetic calorimeter (ECAL), and a brass and scintillator hadron calorimeter (HCAL), each composed of a barrel and two endcap sections. Forward calorimeters extend the pseudorapidity coverage provided by the barrel and endcap detectors. Muons are reconstructed using gas-ionization detectors interleaved with the layers of the steel flux-return yoke outside the solenoid. More detailed descriptions of the CMS detector, together with a definition of the coordinate system used and the relevant kinematic variables, can be found in Refs.~\cite{CMS:2008xjf, CMS:2023gfb}.

Events of interest are selected using a two-tiered trigger system. The first level, composed of custom hardware processors, uses information from the calorimeters and muon detectors to select events at a rate of around 100\unit{kHz}~\cite{CMS:2020cmk}. The second level, known as the high-level trigger, consists of a farm of processors running a version of the full event reconstruction software optimized for fast processing and reduces the event rate to a few kHz before data storage~\cite{CMS:2016ngn,CMS:2024aqx}.

Muons are measured in the pseudorapidity range $\abs{\eta}<2.4$, with detection planes made using three technologies: drift tubes, cathode strip chambers, and resistive-plate chambers. The single muon trigger efficiency exceeds 90\% over the full $\eta$ range, and the efficiency to reconstruct and identify muons is greater than 96\%. Matching muons to tracks measured in the silicon tracker results in a relative transverse momentum resolution, for muons with \pt up to 100\GeV, of 1\% in the barrel and 3\% in the endcaps.

A particle-flow (PF) algorithm~\cite{CMS:2017yfk} is used to reconstruct and identify each individual particle in an event, with an optimized combination of information from the various elements of the CMS detector. The energy of photons is obtained from the ECAL measurement. The energy of an electron is determined from a combination of the electron momentum at the primary interaction vertex as determined by the tracker, the energy of the corresponding ECAL cluster, and the energy sum of all bremsstrahlung photons spatially compatible with originating from the electron track. The energy of a muon is obtained from the curvature of the corresponding track. The energy of a charged hadron is determined from a combination of the momentum measured in the tracker and the matching ECAL and HCAL energy deposits, corrected for the response function of the calorimeters to hadronic showers. Finally, the energy of neutral hadrons is obtained from the corresponding corrected ECAL and HCAL energies.

Jets are reconstructed offline from PF objects using the anti-\kt algorithm~\cite{Cacciari:2008gp,Cacciari:2011ma} with a distance parameter $R=0.4$. Jet momentum is determined as the vectorial sum of all particle momenta in the jet, and is found from simulation to be, on average, within 5 to 10\% of the true momentum over the whole \pt spectrum and detector acceptance. Additional \pp interactions within the same or nearby bunch crossings can contribute extra tracks and calorimetric energy depositions, increasing the apparent jet momentum. To mitigate this effect, tracks identified to be originating from pileup vertices are discarded and an offset correction is applied to correct for remaining contributions~\cite{CMS:2020ebo}. Jet energy corrections are derived from simulation studies so that the average measured energy of jets becomes identical to that of particle-level jets. In situ measurements of the momentum balance in dijet, photon+jet, {\PZ}+jet, and multijet events are used to determine any residual differences between the jet energy scales (JES) in data and in simulation, and appropriate corrections are made~\cite{CMS:2016lmd}. Additional selection criteria are applied to each jet to remove jets potentially dominated by instrumental effects or reconstruction failures~\cite{CMS:2020ebo}.

\section{Monte Carlo simulation}
\label{sec:montecarlo}

The production of \PGU mesons is simulated using the \PYTHIA8.240 event generator~\cite{pythia82}, as well as its newer version \PYTHIA8.310~\cite{pythia83}, which incorporates a novel treatment of quarkonia production in parton showers~\cite{cooke23}. The CP1 and CP5 sets of parameters, tuned by the CMS Collaboration~\cite{cmsTunes} are used in both \PYTHIA versions to model the parton shower and underlying event. The tunes differ in, among other things, the perturbative order at which the running of the strong coupling is evaluated. While CP1 uses the one-loop solution of the renormalization group equation, CP5 uses the corresponding two-loop solution. Additionally, the value of the strong coupling used for the initial- and final-state radiation emissions evaluated at the {\PZ}-pole is $\alpS(m_{\PZ})=0.1365$ for CP1 and $\alpS(m_{\PZ})=0.1180$ for CP5. Since larger values of \alpS translate into larger gluon radiation and, hence, more hadronic activity, it is expected that CP1 will predict less isolated quarkonia, \ie larger hadronic activity in a cone around the \PGU meson. The parton distribution functions (PDF) used for both tunes are also different. While CP5 uses the NNPDF 3.1 NNLO PDF with $\alpS(m_{\PZ})=0.118$~\cite{NNPDF:2017mvq}, the CP1 sample uses the NNPDF 3.1 LO PDF with $\alpS(m_{\PZ})=0.130$~\cite{NNPDF:2017mvq}. The generated samples include \PGUPnS production ($n=1,2,3$) via the colour-singlet~\cite{Bai83, Gas87} and colour-octet mechanisms~\cite{Cho96_1, Cho96_2}. They also account for the feed-down contributions, \ie the production of quarkonia from the decays of heavier states such as \PGcb mesons, $\PGcbDiP{1P}\to\PGUP{1S}\PGg$, where the index $i$ indicates the eigenvalue of the total angular momentum operator $J$. The differential cross section for the production of each \PGUPnS state is calculated following Eq.~\eqref{eq:ldmeSigma}, and the generated \PGUPnS mesons are subsequently forced to decay to a pair of charged muons using the VLL model from the \EVTGEN1.6.0 decay package~\cite{evtgen}, which implements the vector coupling and angular distributions for $\PGUPnS\to\Pell^+\Pell^-$ processes. The response of the CMS detector is simulated using the \GEANTfour package~\cite{geant4}. This detector simulation is only applied to the \PYTHIA8.240 samples.

The detector-level simulated distributions are corrected for several detector effects. First, the four-momenta of simulated jets are smeared to reproduce the jet energy resolution (JER) observed in data~\cite{CMS:2016lmd}. Similarly, the momenta of muons are smeared to match the resolution in data~\cite{rochester}. Additionally, simulated events are reweighted so that the pileup conditions in the simulation reproduce those in data~\cite{pileup}. This is based on the distribution of the average number of \pp interactions per bunch crossing. The muon identification and reconstruction efficiencies measured in simulation are corrected to match those observed in data. These corrections are derived using the tag-and-probe method in $\PZ\to\PGm\PGm$ events~\cite{CMS:2018rym}. An additional correction is needed to account for the finite time resolution of the trigger, which may cause some events to be assigned to an incorrect bunch crossing with a certain probability. To correct for this effect, known as `trigger prefiring'~\cite{CMS:2020cmk, muonTrigger}, dedicated event weights are derived and applied to the simulated samples. These corrections come with associated systematic uncertainties that are described in Section~\ref{sec:systematics}.

Recently, the contributions to \PGU production from the parton shower have been discussed in Ref.~\cite{cooke23} and implemented in \PYTHIA8.310~\cite{pythia83}. In this case, matrix elements for $2\to2$ parton-parton scattering are generated at leading order in QCD, after which the \PYTHIA parton shower is applied. The hadronization of the outgoing partons into \PGU mesons follows analytical expressions for the fragmentation functions of quarks and gluons into the ${}^{2S+1}L_J$ Fock states discussed above. The expressions for such functions are explicitly derived in Refs.~\cite{braaten1, braaten2, braaten3, braaten4, yuan1}. The total cross section is obtained from the partial wave decomposition following Eq.~\eqref{eq:ldmeSigma}.

Event samples for \PGUP{1S}, \PGUP{2S}, and \PGUP{3S} are simulated using \PYTHIA8.310 at the particle level only (\ie after parton shower and hadronization, and without explicit simulation of the detector response), in order to compare with the unfolded data. In this case, the decay of the \PGU resonances to pairs of muons is handled by \PYTHIA itself, and the modelling of the parton shower and the underlying event makes use of both the CP1 and CP5 tunes.

\section{Data set and event reconstruction}
\label{sec:data}

The data set used for the present analysis corresponds to an integrated luminosity of 138\fbinv of \pp collisions at a centre-of-mass energy of 13\TeV, recorded by the CMS experiment in 2016--2018~\cite{LUM-17-003, LUM-17-004, LUM-18-002}.

Events are preselected using dedicated dimuon triggers with different \pt thresholds, ranging from 8 to 12\GeV, depending on the data-taking period. The \PGU candidates are reconstructed as follows: for each event, the muon momentum scale and resolution are corrected using factors derived from \PZ boson samples in order to remove biases of the muon momentum from detector misalignments or magnetic field uncertainties~\cite{rochester}. Muons with $\pt>3\GeV$ are then preselected and required to satisfy basic selection criteria. All oppositely charged muon pairs passing these requirements are fitted to a common vertex, and only those with an invariant mass in the range $7<m_{\PGm\PGm}<13\GeV$ are retained as \PGU candidates. Subsequently, more stringent selection criteria are applied: the \pt and $\eta$ of each muon are required to fulfil $\pt>6\GeV$ and $\abs{\eta}<1.4$. In order to satisfy the trigger requirements, the difference in pseudorapidity between the two muons is required to be $\abs{\Delta\eta}<1.6$. Moreover, the transverse momentum and the rapidity of the dimuon system are required to fulfil $\pt^{\PGm\PGm}>15\GeV$ and $\abs{y^{\PGm\PGm}}<1.2$, respectively. The invariant mass window is narrowed to $8.7<m_{\PGm\PGm}<11.3\GeV$, and both muons are required to spatially match the muon objects passing triggers and to fulfil tighter quality criteria. Finally, if more than one \PGU candidate survives the selection, the one having a smaller $\chi^2$ value for the vertex fit is retained for further analysis.

The selected \PGU candidates are required to be fragmented from a jet. Thus, both muons resulting from the \PGU decay are required to be clustered (\ie to be part of the jet constituents) into one of the jets reconstructed in the event. Since the standard jet energy scale corrections~\cite{CMS:2016lmd} are derived using inclusive jets, which usually do not contain muons, the direct application of these corrections to the jets used in this analysis results in a positive bias of the jet response, with the reconstructed jet \pt being larger, on average, than its true value. To correct for this effect, the four-momenta of the muons is subtracted from the four-momentum of the jet containing them. The remaining four-momentum, which is purely hadronic, is then corrected for its pileup offset, and the response and residual corrections are applied~\cite{CMS:2016lmd}. This approach, the so-called muon subtraction method, reduces the bias in the jet response when evaluated in MC samples with respect to the true jet \pt. Moreover, the method has been tested in data by considering the recoil of a purely hadronic jet (\ie not containing any muons) against a jet containing an \PGU candidate. The bias in the average jet response, defined as the ratio of the \pt of both jets in data, is also reduced. Corrected jets are required to have a \pt above 60\GeV and a pseudorapidity $\abs{\eta} < 1.5$.

\section{Yield determination}
\label{sec:purity}

The \PGU meson selection criteria described in Section~\ref{sec:data} select all candidates in the invariant mass window between 8.7 and 11.3\GeV. Thus, the selected sample contains \PGUP{1S}, \PGUP{2S}, and \PGUP{3S} candidates, as well as backgrounds arising from muon misidentification or low-mass Drell--Yan processes. The sample is divided in three \pt bins with lower boundaries at 60, 75, and 100\GeV, and each of these is further binned into different $z$ and \ptrel bins. To disentangle the four different components of the sample for each of these bins, maximum likelihood fits are performed to the invariant mass distribution of the dimuon pairs forming the \PGU candidates. The fraction of each \PGUPnS species in a particular phase space region can be determined by means of a binned maximum likelihood fit to the invariant mass distribution of the dimuon candidates. The fit model can be expressed as
\begin{equation}\label{eq:fitModel}
    \mathcal{F}(m) = \sum_n\lambda_n\mathcal{F}_n(m)+\lambda_{\mathrm{B}}\mathcal{F}_{\mathrm{B}}(m),
\end{equation}
where $\mathcal{F}_n$ is the probability density function (pdf) for the \PGUPnS component and $\mathcal{F}_{\mathrm{B}}$ is the pdf for the background component. The coefficients $\lambda_n$ and $\lambda_{\mathrm{B}}$ are the relative fractions of each component and are required to satisfy the closure relation $\sum_i\lambda_i=1$.
The pdfs of the \PGUPnS resonances are described by a Crystal-Ball function~\cite{Oreglia:1980cs,osti_6237411}, \ie
\begin{equation}\label{eq:crystalBall}
    \mathcal{F}_n\big(m\big\vert\mu_n,\sigma_n,\nu,\alpha\big) = \begin{cases}
        N\exp\left[-\dfrac{(x-\mu_n)^2}{2\sigma_n^2}\right] & \text{if } \dfrac{x-\mu_n}{\sigma_n}>-\alpha, \\[3ex]
        NA\left(B-\frac{x-\mu_n}{\sigma_n}\right)^{-\nu} & \text{if } \dfrac{x-\mu_n}{\sigma_n}\leq-\alpha,
    \end{cases}
\end{equation}
where $N$ is a normalization factor and $A$ and $B$ are given by
\begin{equation}
    A = \left(\frac{\nu}{\abs{\alpha}}\right)^\nu\exp\left[-\frac{\abs{\alpha}^2}{2}\right];
    \qquad
    B = \frac{\nu}{\abs{\alpha}}-\alpha.
\end{equation}
The parameters $\mu_n$ and $\sigma_n$ define the mean and standard deviation from the Gaussian core distributions, while $\nu$ and $\alpha$ parameterize the shape of the low-mass tails, arising primarily from the energy loss induced by photon bremsstrahlung. The background component is described by a third-order Chebyshev polynomial, \ie
\begin{equation*}
    \mathcal{F}_{\mathrm{B}}\big(m\big\vert p_1,p_2,p_3\big) = 1+p_1T_1\big(x(m)\big)+p_2T_2\big(x(m)\big)+p_3T_3\big(x(m)\big),
\end{equation*}
where $T_1(x)=x$, $T_2(x)=2x^2-1$, $T_3(x)=4x^3-3x$, and $x(m)$ is a linear transformation that maps the mass interval $[8.7, 11.3]\GeV$ onto $x\in[-1,1]$. This is the lowest order polynomial that is able to provide a good fit in all regions of the phase space. Using fourth-order polynomials instead produces yield results which differ from the central result by less than 0.1\%.

The free parameters for each of the functions are constrained so that each component is normalised to unit area, \ie $\int\mathcal{F}_i(m)\rd m=1$ in the mass window between 8.7 and 11.3\GeV. Moreover, since the masses of the \PGU resonances are well known~\cite{pdg}, the values of $\mu_n$ and $\sigma_n$ in Eq.~\eqref{eq:crystalBall} are constrained as
\begin{equation*}
    \frac{\mu_n}{\mu_1} = \frac{\sigma_n}{\sigma_1} = \left[\frac{M_{\PGUPnS}}{M_{\PGUP{1S}}}\right]_{\text{PDG}},
\end{equation*}
where PDG denotes the world-average \PGUPnS masses~\cite{pdg}. Additionally, the parameters $n$ and $\alpha$ are set to be the same for all resonant contributions. They are fitted from the inclusive \PGUPnS sample (\ie without any kinematic restrictions) and used everywhere across the phase space of the analysis.

\begin{figure}[!ht]
\includegraphics[width=0.49\textwidth]{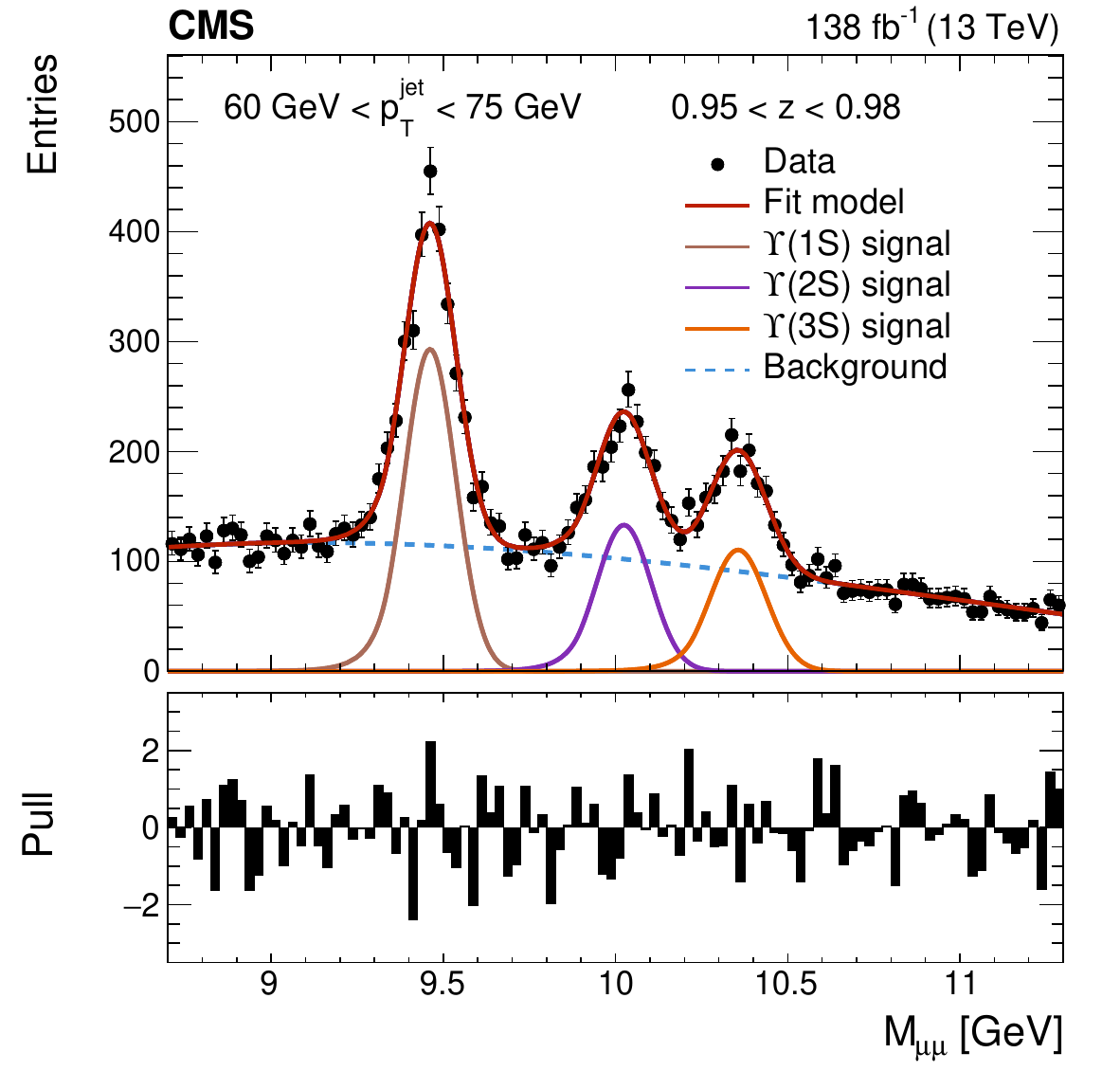}%
\hfill%
\includegraphics[width=0.49\textwidth]{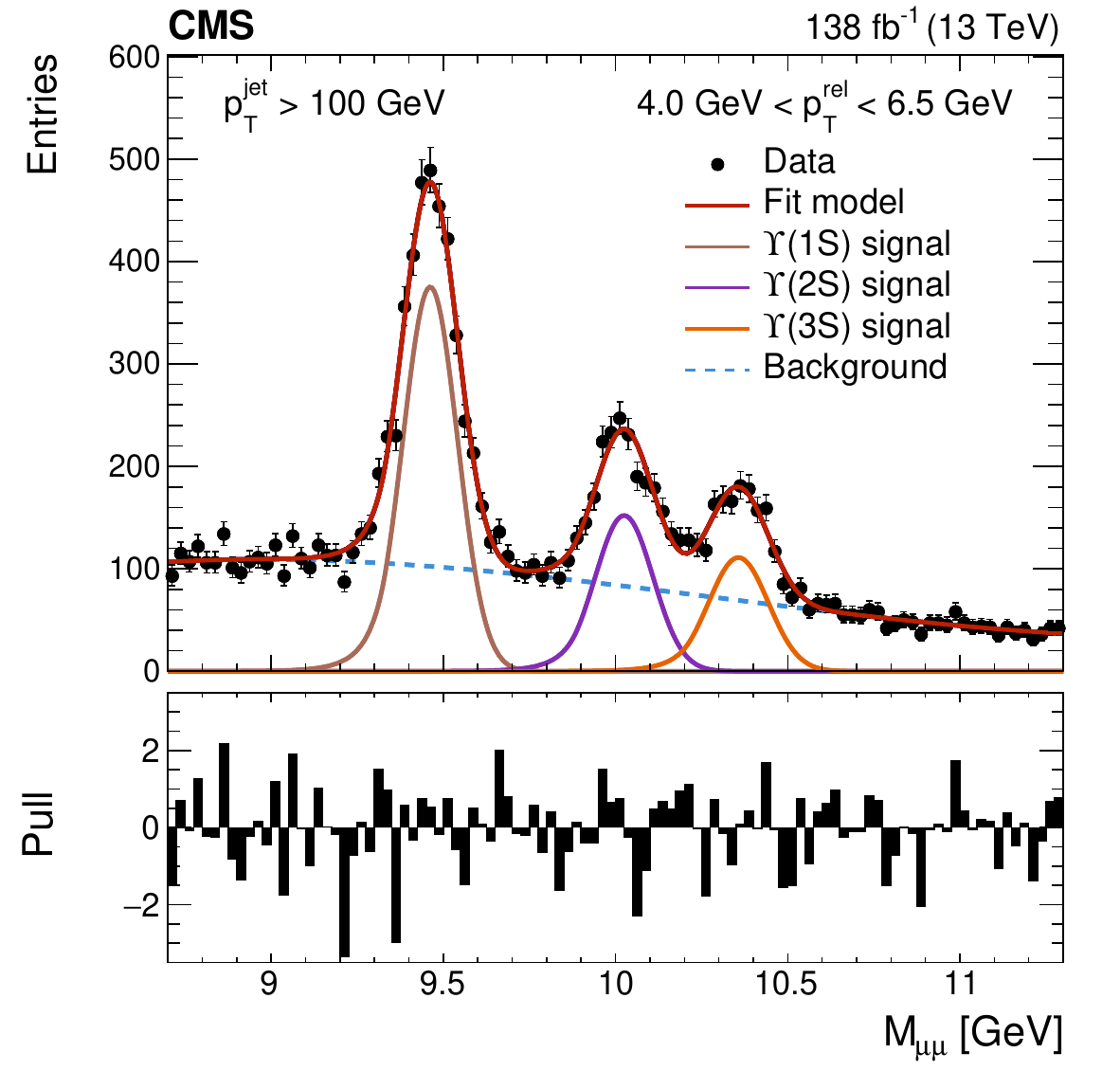}
\caption{Fits to the invariant mass distributions of \PGUPnS candidates in two representative ranges of the jet \pt and the fragmentation variables $z$ (left) and \ptrel (right). The fit components, including signal and background functions, are also shown on the upper panels. The vertical error bars represent the statistical uncertainties on the data points. The lower panel shows the difference between the data and the fitted function, divided by the statistical uncertainty of the data. The model shows, in general, a good agreement with the data.}
\label{fig:massFits}
\end{figure}

Figure~\ref{fig:massFits} shows the results of the invariant mass fits for two representative $(\ptjet, z)$ and $(\ptjet, \ptrel)$ bins. In order to quantify the agreement, the lower panels show the pulls of the data with respect to the fit model, \ie the difference between the data and the model divided by the statistical error of the data points. As can be seen from the figure, the fitted functions from Eq.~\eqref{eq:fitModel} provide a good description of the dimuon invariant mass distributions across all regions of the phase space, with most of the experimental points falling within one standard deviation of the fit model.

\begin{figure}[!ht]
\includegraphics[width=0.49\textwidth]{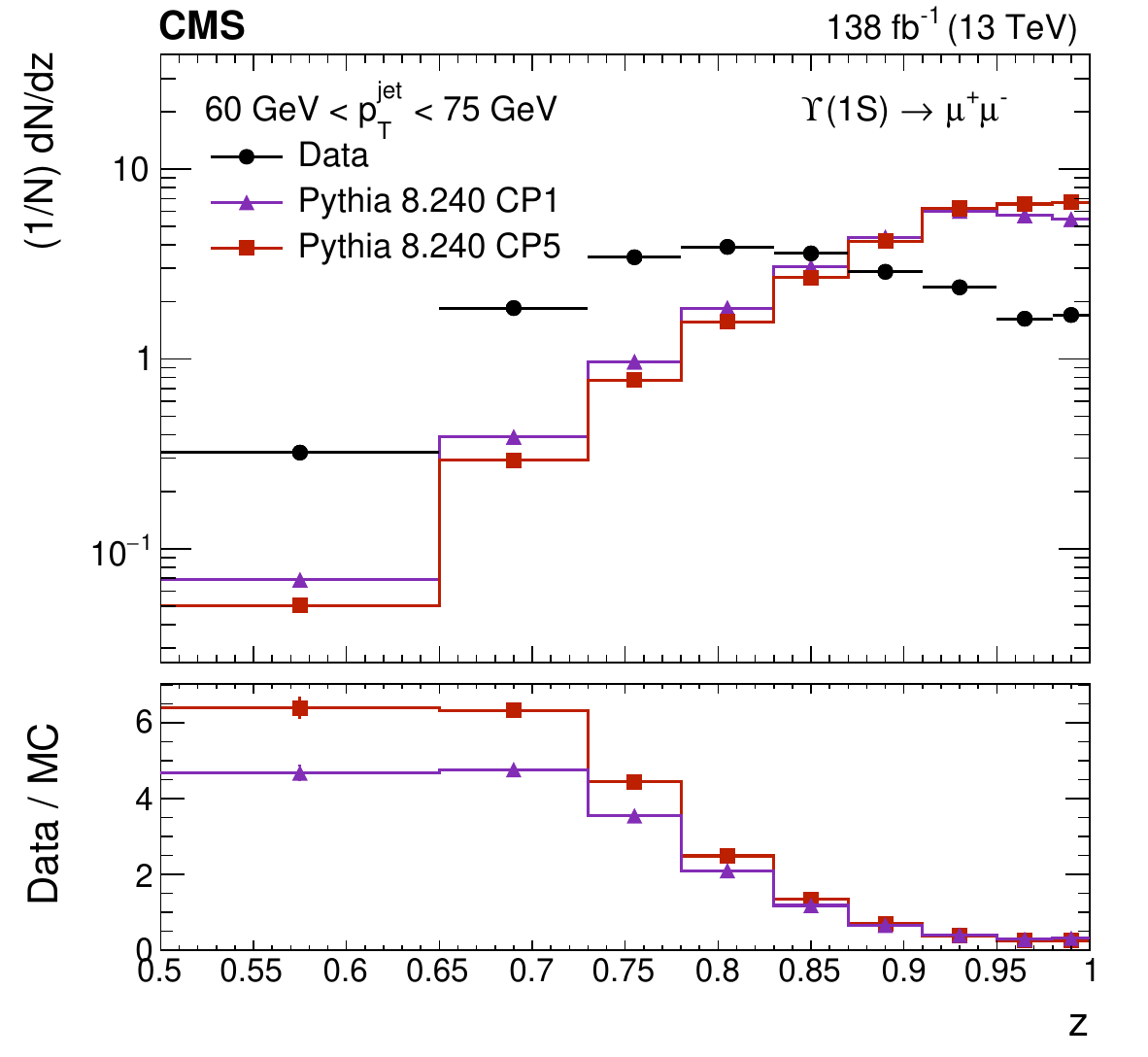}%
\hfill%
\includegraphics[width=0.49\textwidth]{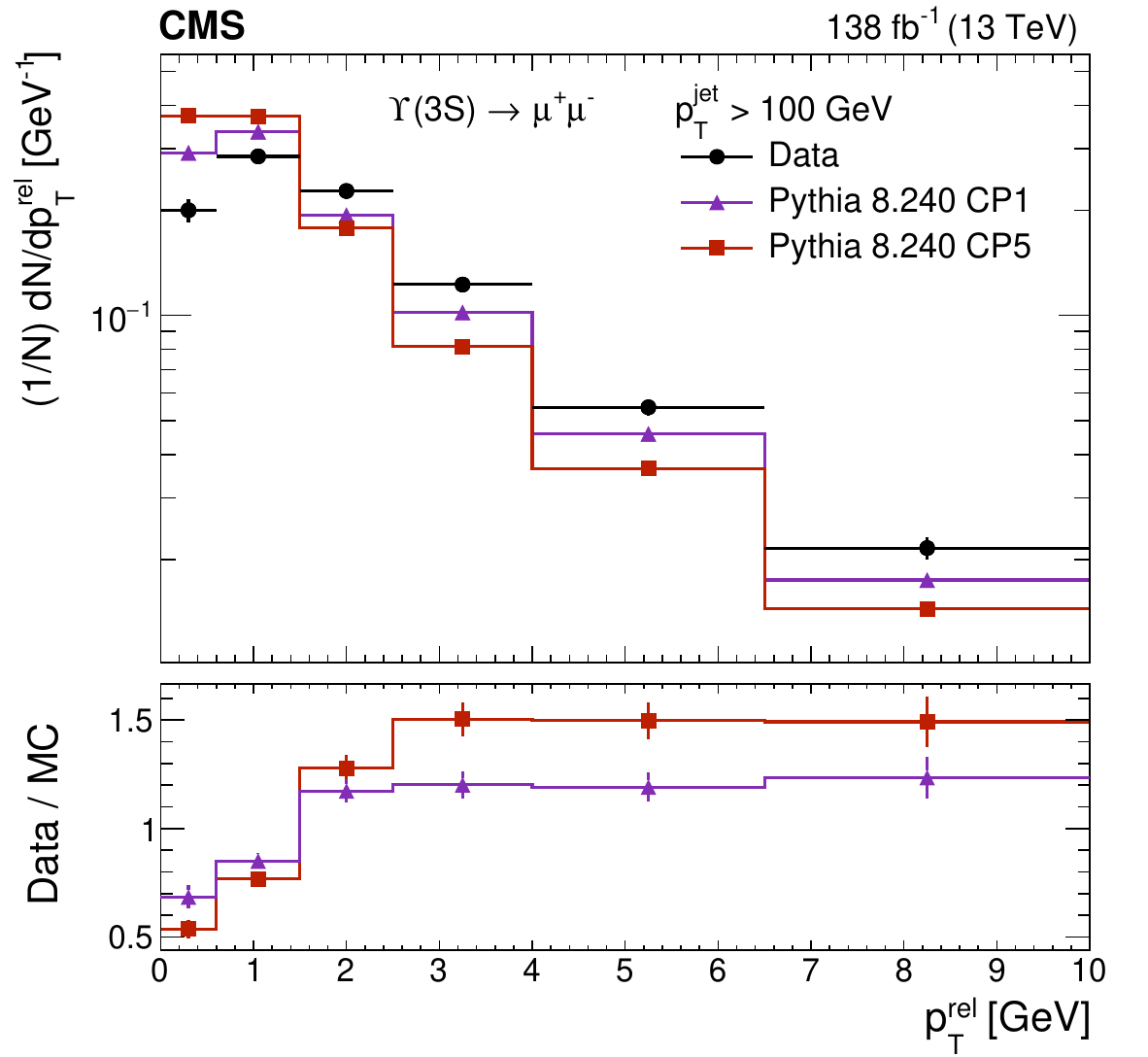}
\caption{Detector-level longitudinal profile for \PGUP{1S} mesons (left) and transverse profile for \PGUP{3S} mesons (right) for two different ranges of the jet transverse momentum. The error bars on the data distributions represent the statistical uncertainties.}
\label{fig:detLevel}
\end{figure}

Once the yields of each \PGUPnS species are determined using these fits, the $z$ and \ptrel distributions can be directly compared to MC predictions at the detector level. Figure~\ref{fig:detLevel} shows the detector-level $z$ and \ptrel results for \PGUP{1S} and \PGUP{3S} in two particular bins of the jet \pt, compared with the MC expectations from \PYTHIA8.240 with the CP1 and CP5 tunes. Significant differences between the data and the simulations are observed. Namely, \PYTHIA predicts \PGUPnS mesons to be produced in association with less hadronic activity (and with a smaller momentum with respect to the jet) than what is observed in data. This effect, observed for the three \PGUPnS mesons, is similar to what was observed in previous measurements by LHCb~\cite{lhcbJpsiFrag} and CMS~\cite{cmsJpsiFrag} for the \PJGy meson.

\section{Unfolding}
\label{sec:unfolding}

The detector-level longitudinal and transverse profiles of \PGUP{1S}, \PGUP{2S}, and \PGUP{3S} mesons are corrected for detector effects, such as inefficiencies or migrations between different bins caused by the finite resolution of the detector.

In the simulation, the definition of the particle-level phase space closely follows the kinematic selection described in Section~\ref{sec:data} at the detector level. First, events are required to contain exactly one \PGUPnS meson, with $n=1$, 2, or 3, decaying into a pair of opposite-sign muons. These two muons must fulfil $\pt>6\GeV$, while the dimuon system is required to have a transverse momentum $\pt^{\PGm\PGm}>15\GeV$ and a rapidity $\abs{y^{\PGm\PGm}}<1.2$. Particle-level jets are clustered using the anti-\kt algorithm with $R=0.4$. Charged leptons and photons are included in the clustering, whereas neutrinos are explicitly excluded from the jet inputs. Jets are required to have $\pt>60\GeV$ and $\abs{\eta}<1.5$. Finally, the muons from the \PGU meson decay are required to be part of the jet constituents to ensure that the energy of the \PGU meson is fully contained by the jet.

The correction of the measured distributions to particle level needs to account for the resolution effects, as well as for the detector inefficiencies. To this end, it is customary to consider a linear problem of the form
\begin{equation}\label{eq:unfolding}
    R_i = \sum_{j=1}^N\frac{\mathcal{E}_j}{\mathcal{P}_i}A_{ij}T_j,
\end{equation}
where $R_i$ is the value of the detector-level distribution at bin $i$ and $T_j$ is the equivalent at the particle level. While the matrix $A_{ij}$ parameterizes the migrations among bins caused by the detector resolution effects, the factors $\mathcal{E}_j$ and $\mathcal{P}_i$, respectively, parameterize the efficiency, the fraction of generated events that are actually reconstructed by the detector, and the purity, the fraction of the reconstructed events that have an actual counterpart at particle level. All these corrections are obtained using the \PYTHIA8.240 CP1 sample.

\begin{figure}[!ht]
\includegraphics[width=0.49\textwidth]{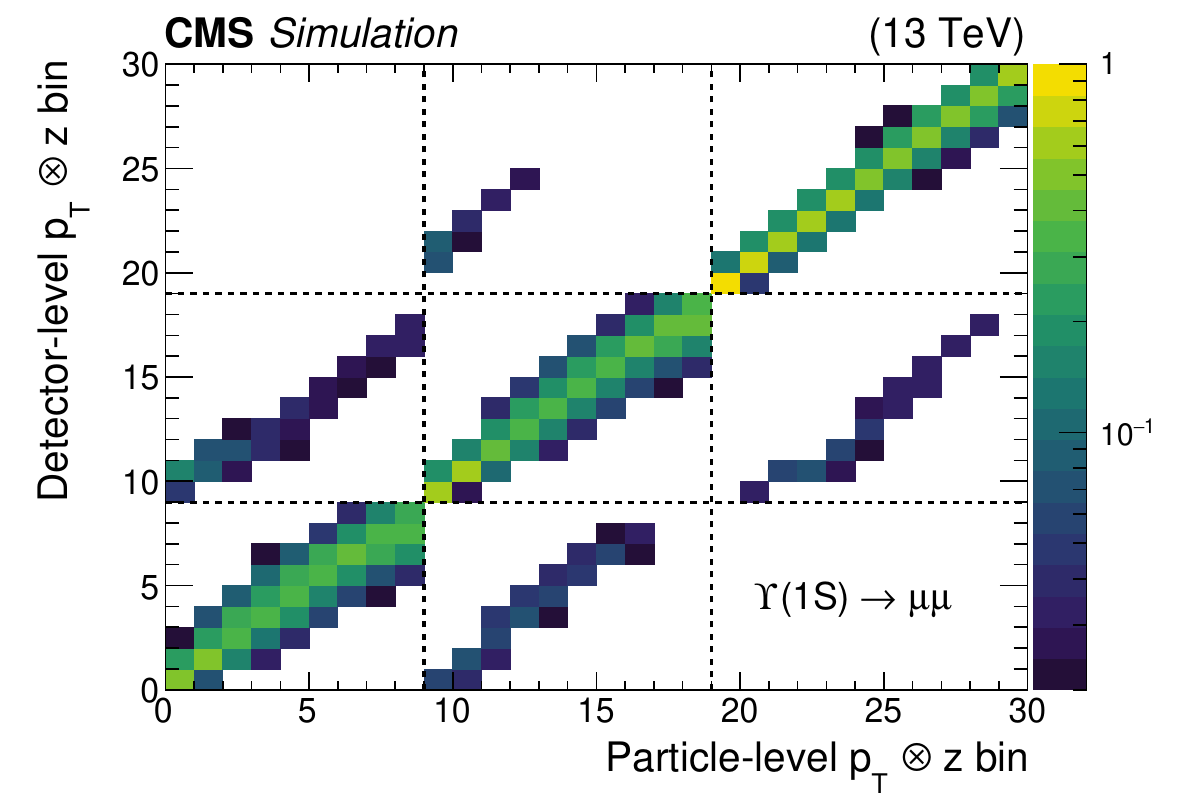}%
\hfill%
\includegraphics[width=0.49\textwidth]{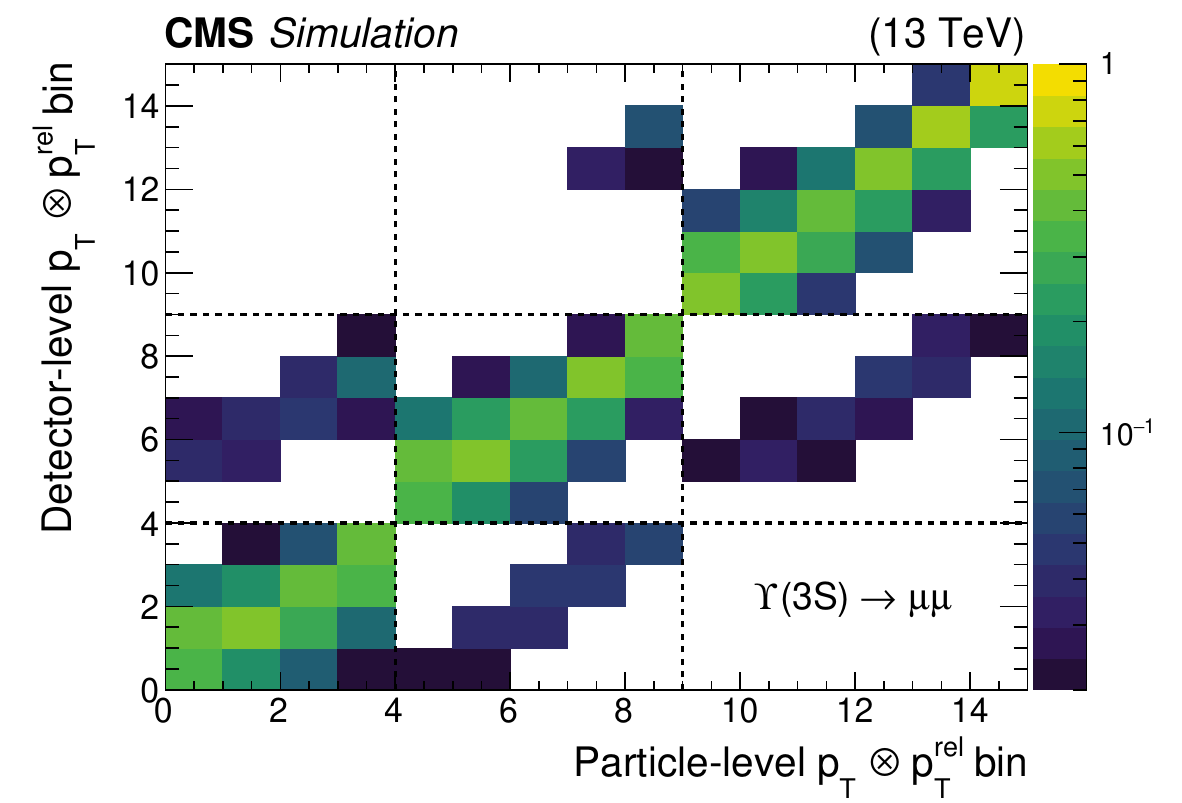}
\caption{Transfer matrices for the longitudinal profile of jets containing \PGUP{1S} mesons (left) and for the transverse profile of jets containing \PGUP{3S} (right). Each transfer matrix parameterizes the migrations among the three bins on the jet \pt used in the measurement, shown as boxes delimited by dashed lines, as well as among bins on $z$ or \ptrel.}
\label{fig:matrix}
\end{figure}

The transfer matrices $A_{ij}$ are obtained separately for jets containing \PGUP{1S}, \PGUP{2S}, and \PGUP{3S} mesons, and contain information on the migrations among jet \pt bins, as well as among $z$ or \ptrel bins. Figure~\ref{fig:matrix} shows the transfer matrices for the longitudinal profile of jets containing \PGUP{1S} and the transverse profile for jets containing \PGUP{3S} mesons. Each of the large boxes represents a jet \pt bin, while the migrations between $z$ (\ptrel) bins are parameterized by the smaller squares. In general, the resolution on $z$ is better than that on \ptrel because of the larger sensitivity of the latter to the angle between the directions of the jet and the quarkonium. The binning of the measurements is chosen to minimize migration effects, while keeping the number of bins as large as possible.

While, in principle, it may be possible to directly invert the matrix $A_{ij}$ in Eq.~\eqref{eq:unfolding}, in practice this can pose several problems such as very large statistical fluctuations or even negative values of the differential cross section under study. A way to circumvent these potential issues is to use a regularized unfolding method and, in particular, the so-called D'Agostini method~\cite{dagostini} as implemented in the \textsc{RooUnfold} package~\cite{roounfold}. This method provides an iterative algorithm for finding the maximum likelihood estimator of the unfolding problem. The method naturally includes a free parameter, the number of iterations, which acts as a regularization parameter on the unfolding. This parameter is chosen based on the differences between the unfolded distributions for iteration $n+1$ and the previous one, $n$. These differences are evaluated using a $\chi^2$ test which takes into account all correlations between the bins of the measurement. The smallest value of $n$ at which the $p$-value associated to the $\chi^2$ is larger than 0.05 is taken as the optimal number of iterations. The method is validated in MC by evaluating the closure between the unfolded detector-level distribution and the particle-level distribution of two statistically independent samples.

The correlations between the statistical uncertainties among the bins of the measurement, as well as those between the particle-level and the detector-level distributions, are estimated using Poisson-distributed pseudo-experiments. The precise knowledge of these correlations allows for a better estimation of the statistical uncertainties.

\section{Systematic uncertainties}
\label{sec:systematics}

The systematic uncertainties in the measurement can be divided into three categories. The first comprises the systematic uncertainties affecting the object reconstruction, such as the uncertainties in the calibration of the jet four-momentum using the JES corrections, the JER, or the muon efficiency corrections. The second category contains uncertainties in other aspects of the event reconstruction, such as the pileup modelling or the prefiring effect in the muon trigger described in Section~\ref{sec:montecarlo}. The third category contains the uncertainty in the correction for detector effects, arising from the model dependence of the unfolding procedure described in Section~\ref{sec:unfolding}. The uncertainties are propagated through the unfolding procedure by varying the transfer matrices as well as the efficiency and purity corrections in Eq.~\eqref{eq:unfolding}. The total systematic uncertainties are obtained as the sum in quadrature of all these sources.

\subsection{Object reconstruction uncertainties}
\begin{itemize}
\item \textit{Jet energy scale corrections:} The jet energy scale uncertainty is propagated by shifting the \pt of the detector-level simulated jets, after applying the muon subtraction described in Section~\ref{sec:data}, according to the $(\pt,\eta)$ dependent JES uncertainties~\cite{CMS:2016lmd}. The JES uncertainties include 27 independent nuisance parameters, including uncertainties in the in situ calibrations, the intercalibration, flavour-dependent uncertainties, as well as other uncorrelated sources. The impact of the JES uncertainties is up to 7\% on the measured distributions.
{\tolerance=800
\item \textit{Jet energy resolution:} The four-momenta of muon-subtracted, simulated jets are smeared to reproduce the JER as measured in data~\cite{CMS:2016lmd}. This correction has an associated uncertainty of 2 to 4\%, depending on the jet transverse momentum, which is propagated to the measured distributions. Its impact on the measurement is below 1\%.\par}
\item \textit{Muon efficiency:} The muon efficiency is corrected in the simulation by means of scale factors depending on the transverse momentum and pseudorapidity of the muons~\cite{cmsWZ_7TeV}. The uncertainties in these scale factors~\cite{CMS:2018rym} are propagated to the measured distributions through the unfolding procedure. Its impact on the measurement is between 1 and 5\%.
\item \textit{Muon momentum scale:} The uncertainty in the muon momentum scale~\cite{rochester} is also propagated to the measured distributions. Its impact is found to be between 1 and 5\%.
\end{itemize}

\subsection{Event reconstruction uncertainties}
\begin{itemize}
\item \textit{Pileup:} The uncertainty in the pileup reweighting procedure described in Section~\ref{sec:montecarlo} is estimated by considering a 4.6\% uncertainty on the \pp inelastic cross section~\cite{pileup}, which is propagated to the measured distributions through the unfolding procedure. Its impact on the measurement is below 3\%.
\item \textit{Prefiring:} The uncertainty in the prefiring corrections discussed in Section~\ref{sec:montecarlo} is estimated by considering 20\% shifts to the prefiring probability of each physics object. The effect of these shifts is propagated through the unfolding procedure, yielding a total impact of about 0.1\%.
\end{itemize}

\subsection{Unfolding uncertainties}
\begin{figure}[!ht]
\includegraphics[width=0.48\textwidth]{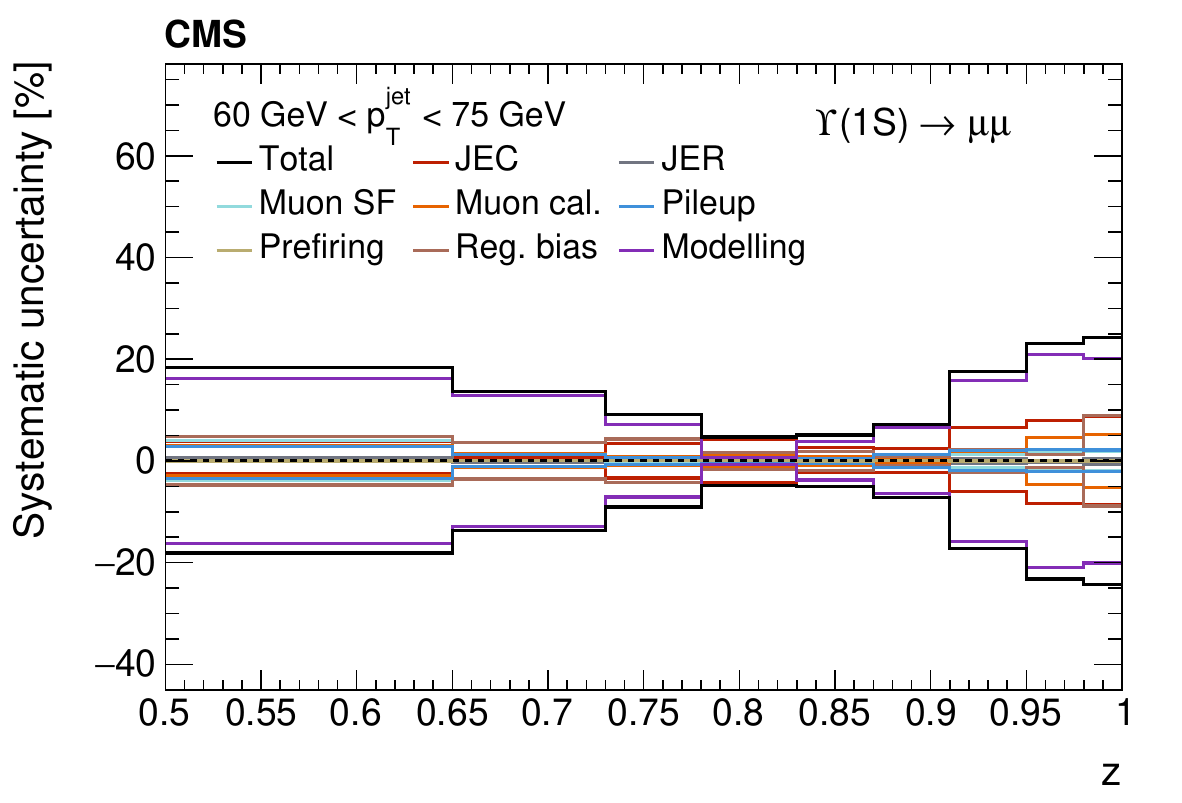}%
\hfill%
\includegraphics[width=0.48\textwidth]{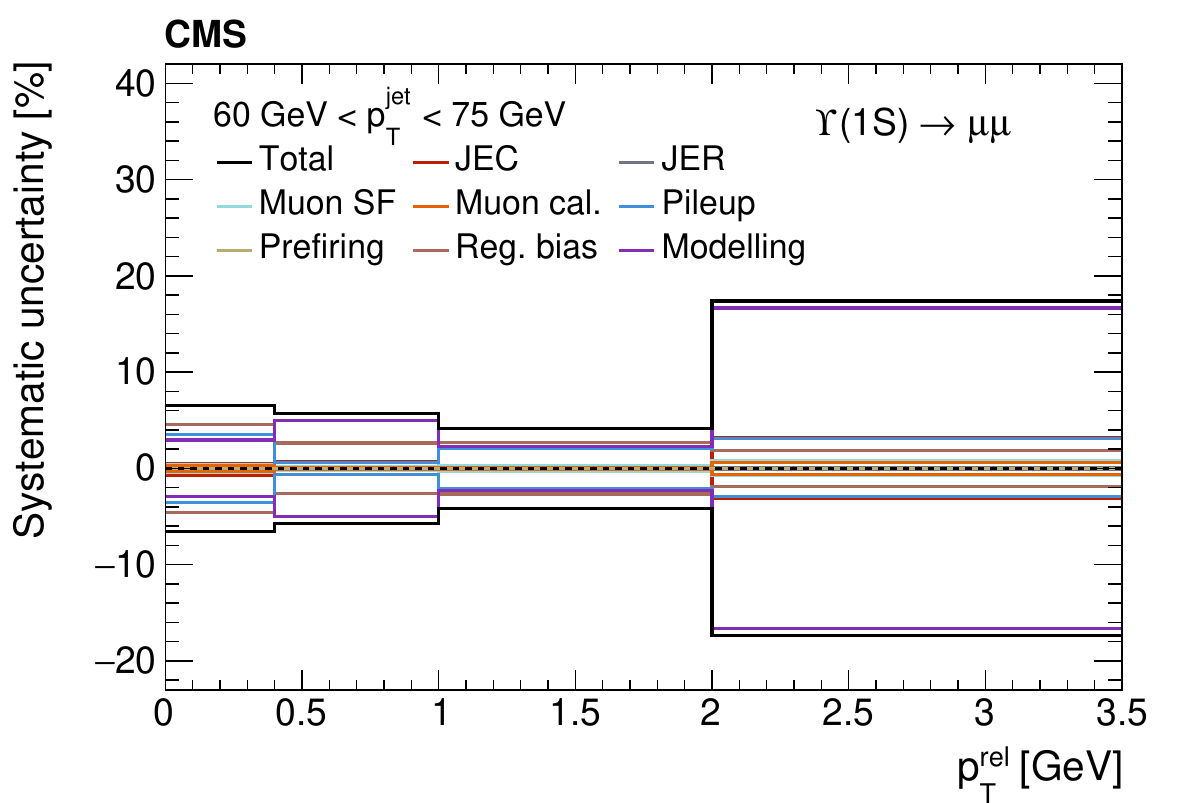} \\
\includegraphics[width=0.48\textwidth]{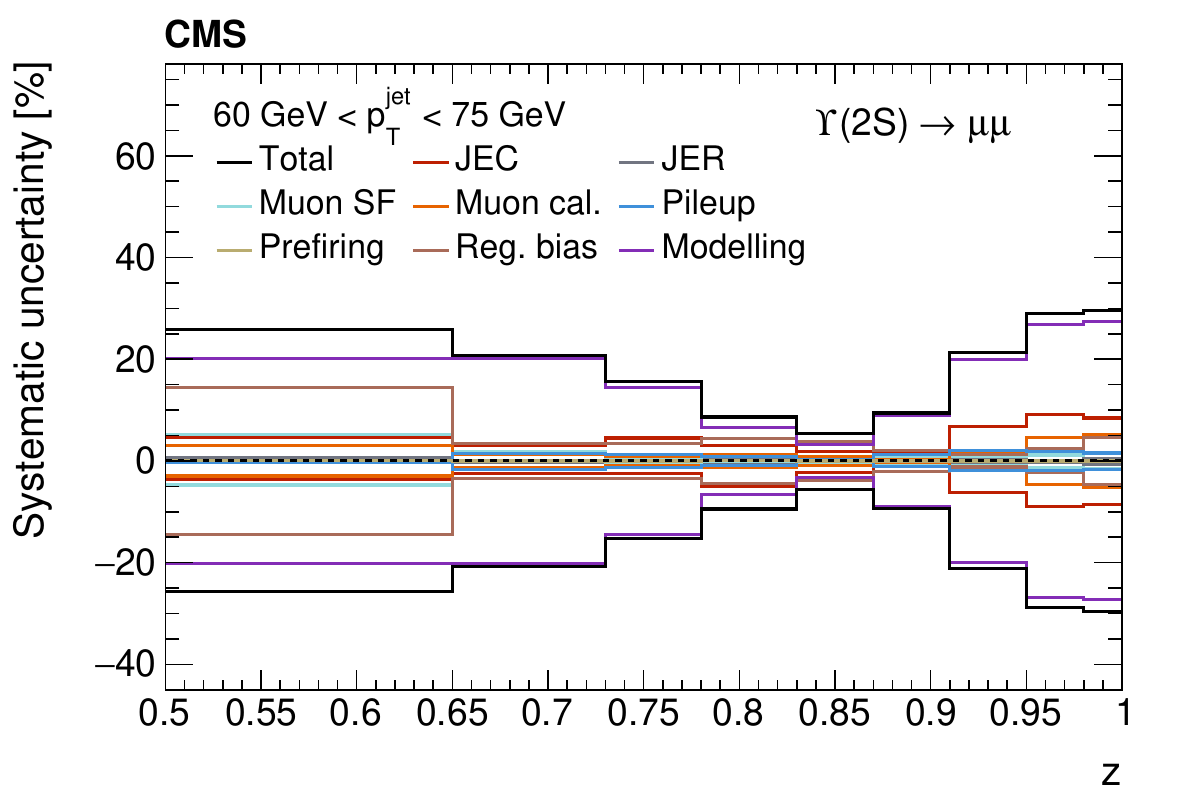}%
\hfill%
\includegraphics[width=0.48\textwidth]{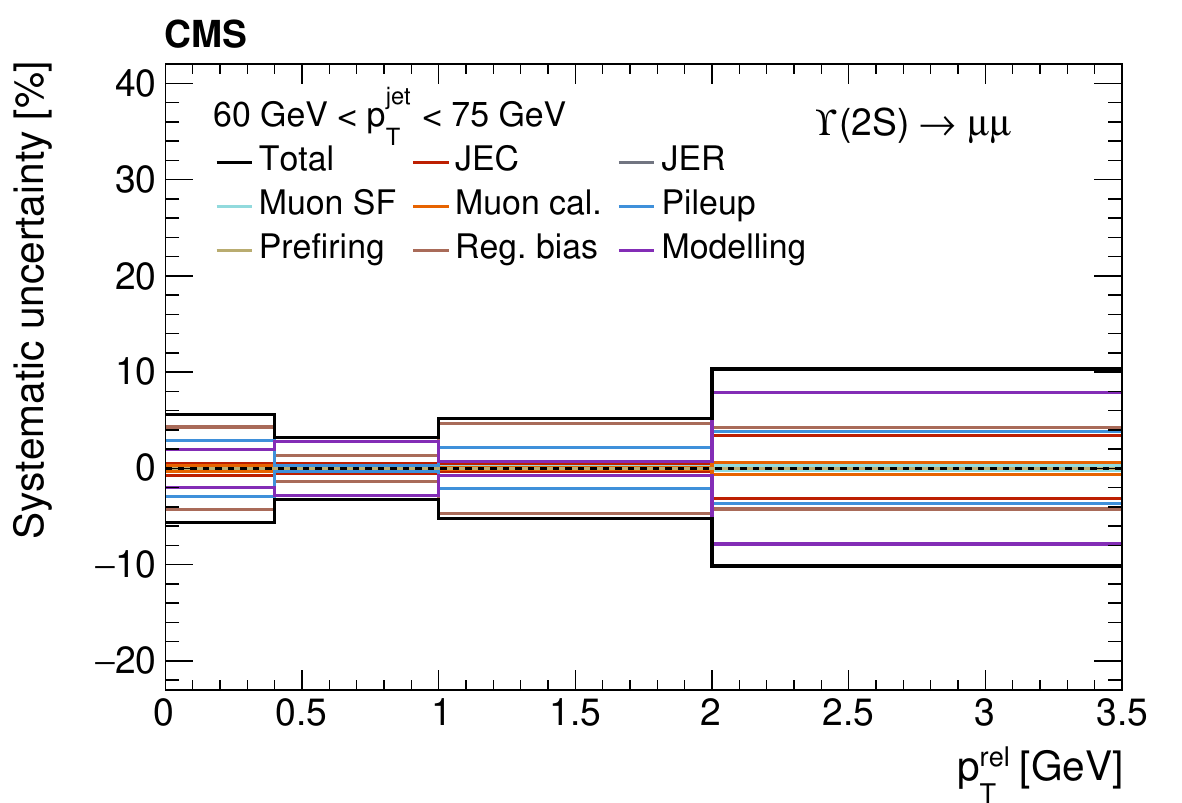} \\
\includegraphics[width=0.48\textwidth]{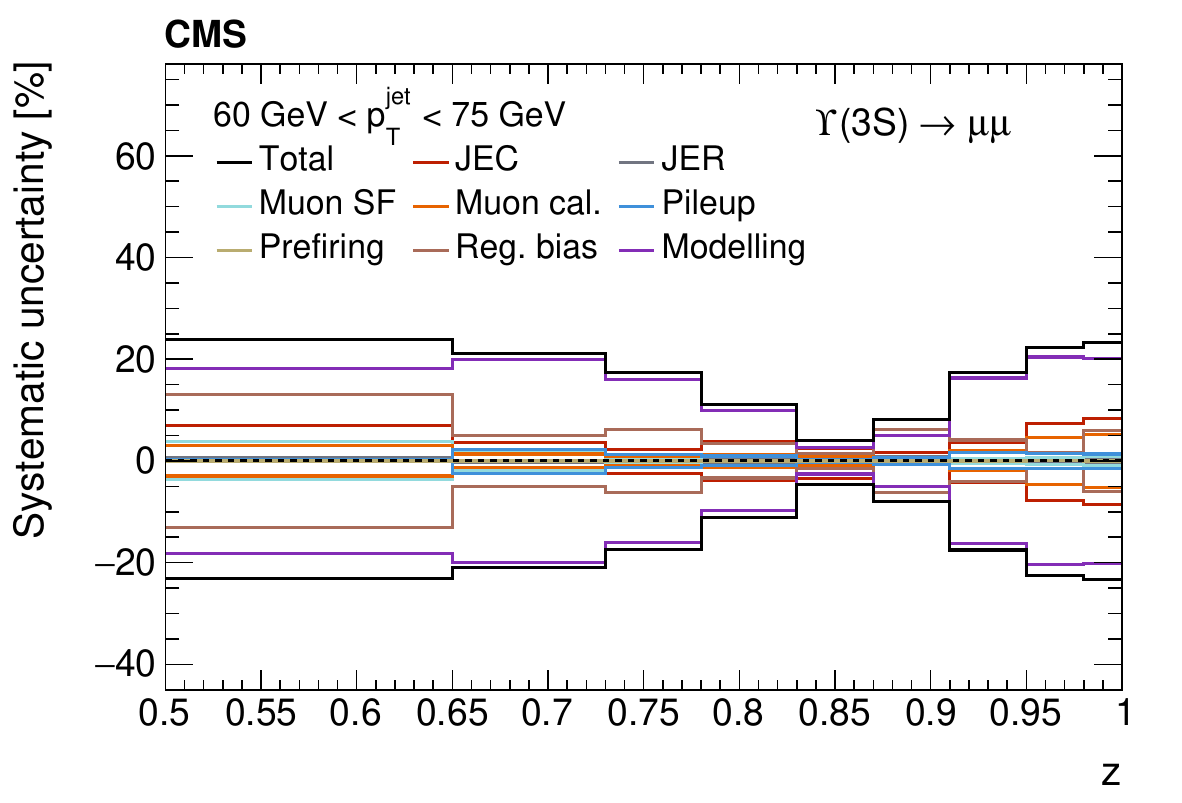}%
\hfill%
\includegraphics[width=0.48\textwidth]{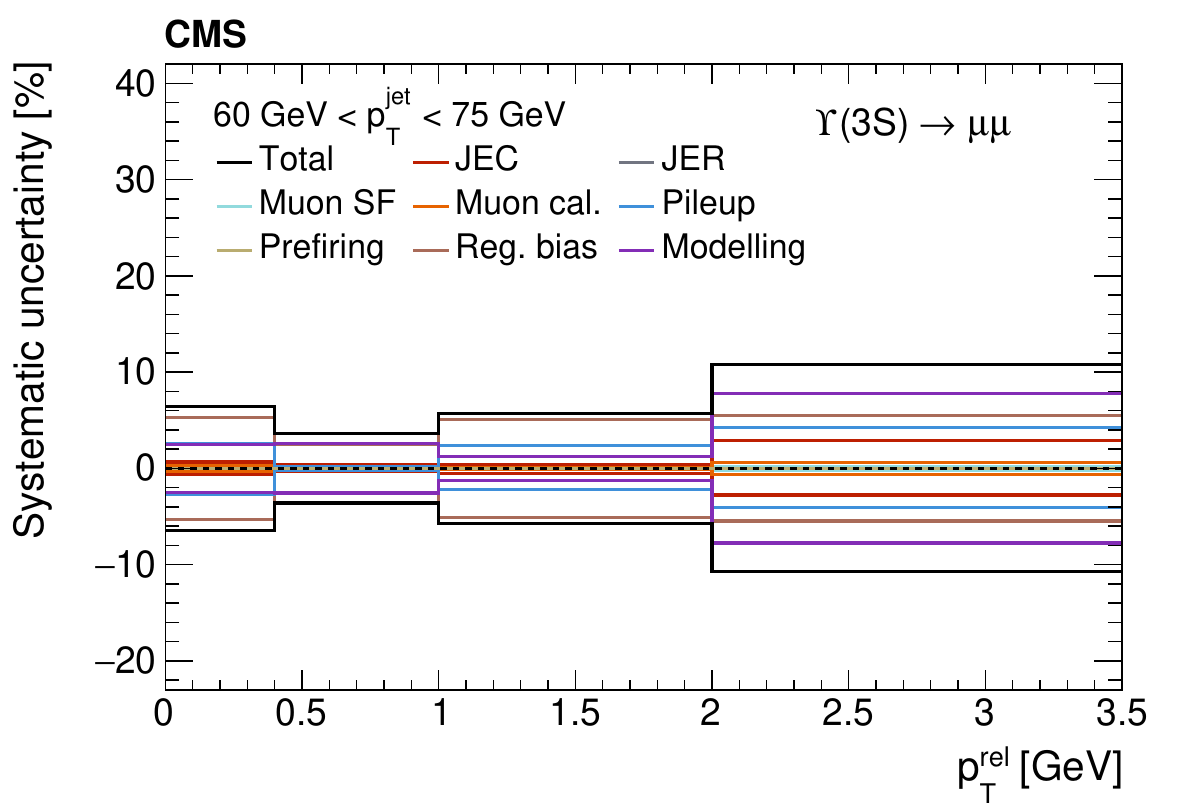}
\caption{Relative values of the systematic uncertainties discussed in the text, together with the total uncertainty band, for the longitudinal and transverse profiles of \PGUP{1S} (top), \PGUP{2S} (middle) and \PGUP{3S} mesons (bottom) in jets with $60<\pt<75\GeV$.}
\label{fig:systematics}
\end{figure}

\begin{itemize}
\item \textit{Regularization bias:} The bias intrinsic to the D'Agostini method is estimated by considering an alternative MC model, in this case \PYTHIA8.240 with the CP5 tune, to define the regularization and treating the \PYTHIA8.240 CP1 sample as pseudo-data. The result is compared with the particle-level distribution of the latter. The uncertainty is defined as the difference between the two, and its impact is between 2 and 15\% for $z$ and typically below 5\% for \ptrel.
\item \textit{Model uncertainty:} The uncertainty due to the inaccurate description of the data by the MC simulation used for the unfolding is estimated by reweighting the \PYTHIA8.240 CP1 sample so that the detector-level distributions of the jet and the \PGUPnS transverse momenta are matched in data and simulation. The reweighted sample is then used to perform an alternative unfolding, which is compared with the central result. The differences between the two define the model uncertainty, which has a typical size between 10 and 20\%, but can reach peak values over 50\% in the collinear region $z\approx1$, where the sensitivity to non-perturbative effects becomes important~\cite{Chapon:2020heu}. This is the dominant uncertainty in the analysis.
\end{itemize}
The relative values obtained for all systematic sources, together with the total uncertainty band, are shown in Fig.~\ref{fig:systematics} for the longitudinal and the transverse profiles for \PGUP{1S}, \PGUP{2S} and \PGUP{3S} mesons in jets with transverse momentum between 60 and 75\GeV.

Theoretical uncertainties associated with the renormalization and factorization scales entering the perturbative calculation of the cross section, as well as with the PDF, are not evaluated for the MC predictions used in this measurement. Since the matrix element is calculated at leading order, variations of the renormalization and factorization scales by a factor of two are not expected to provide a reliable estimate of the uncertainty due to missing higher-order corrections. Furthermore, the measured cross sections are normalized to unit area, so their sensitivity to the PDF is strongly suppressed and the distributions depend primarily on the details of fragmentation.

\section{Results}
\label{sec:results}

The results at the particle level, obtained from the unfolding procedure described in Section~\ref{sec:unfolding}, are presented in this section, together with the systematic uncertainties described in Section~\ref{sec:systematics}. The distributions are normalized to unit area for each interval of the jet \pt under consideration. The values are compared with the predictions from \PYTHIA8.240 and \PYTHIA8.310 with the CP1 and CP5 tunes. In the former \PYTHIA version, \PGUPnS production is modelled by NRQCD matrix elements for colour-singlet and colour-octet mechanisms in the hard process~\cite{pythia82} whereas, for the latter, NRQCD fragmentation processes are implemented into time-like parton showers~\cite{cooke23}.

\begin{figure}[!b]
\includegraphics[width=0.48\textwidth]{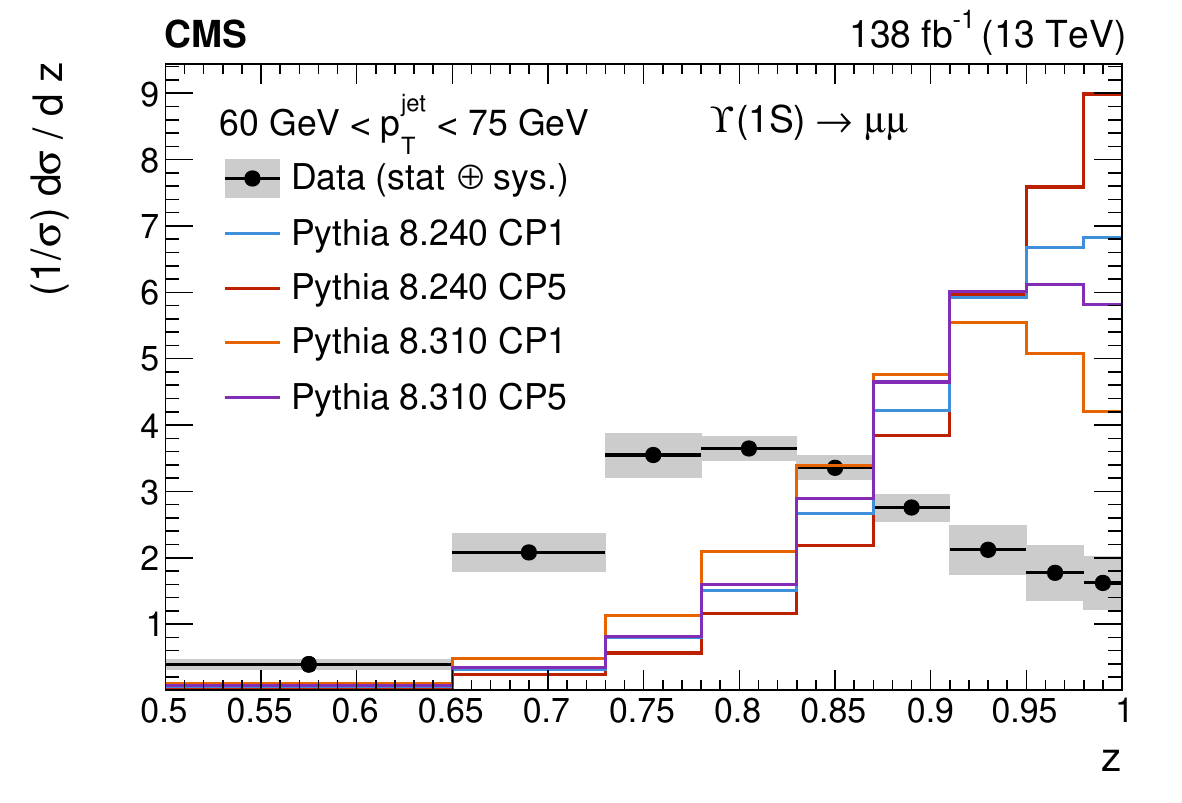}%
\hfill%
\includegraphics[width=0.48\textwidth]{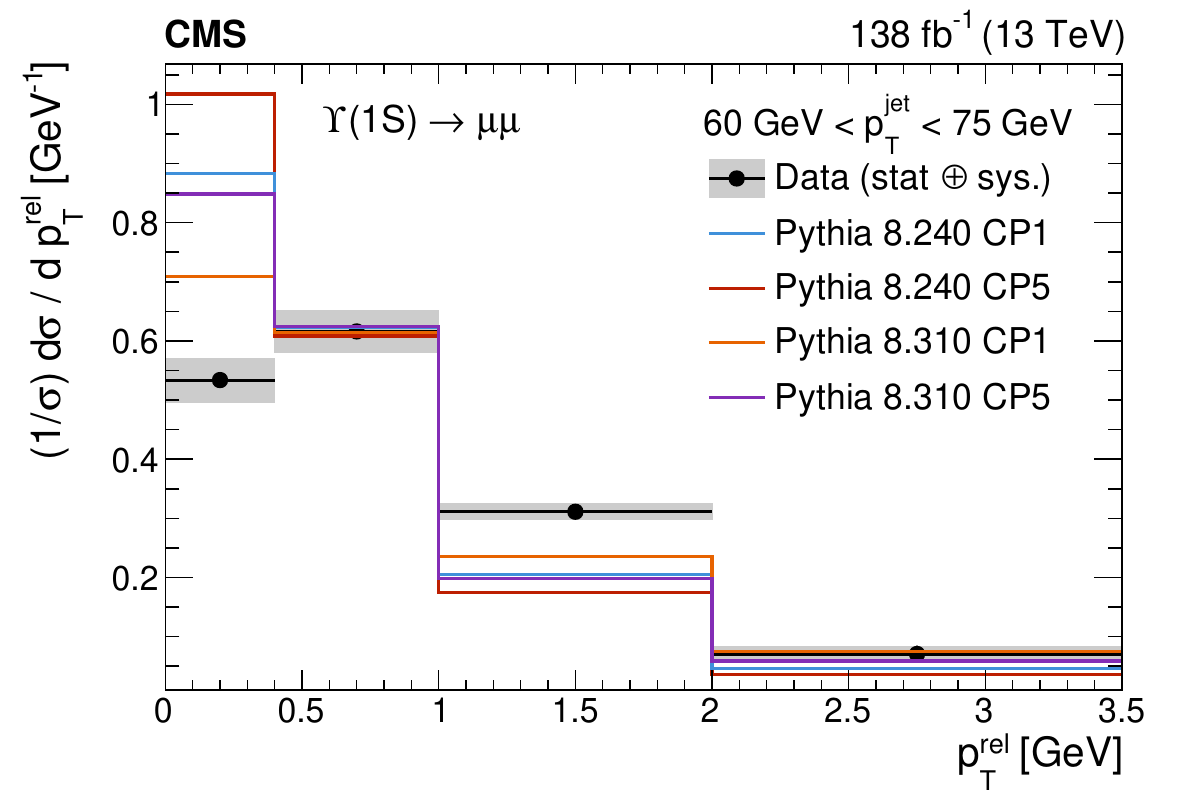} \\
\includegraphics[width=0.48\textwidth]{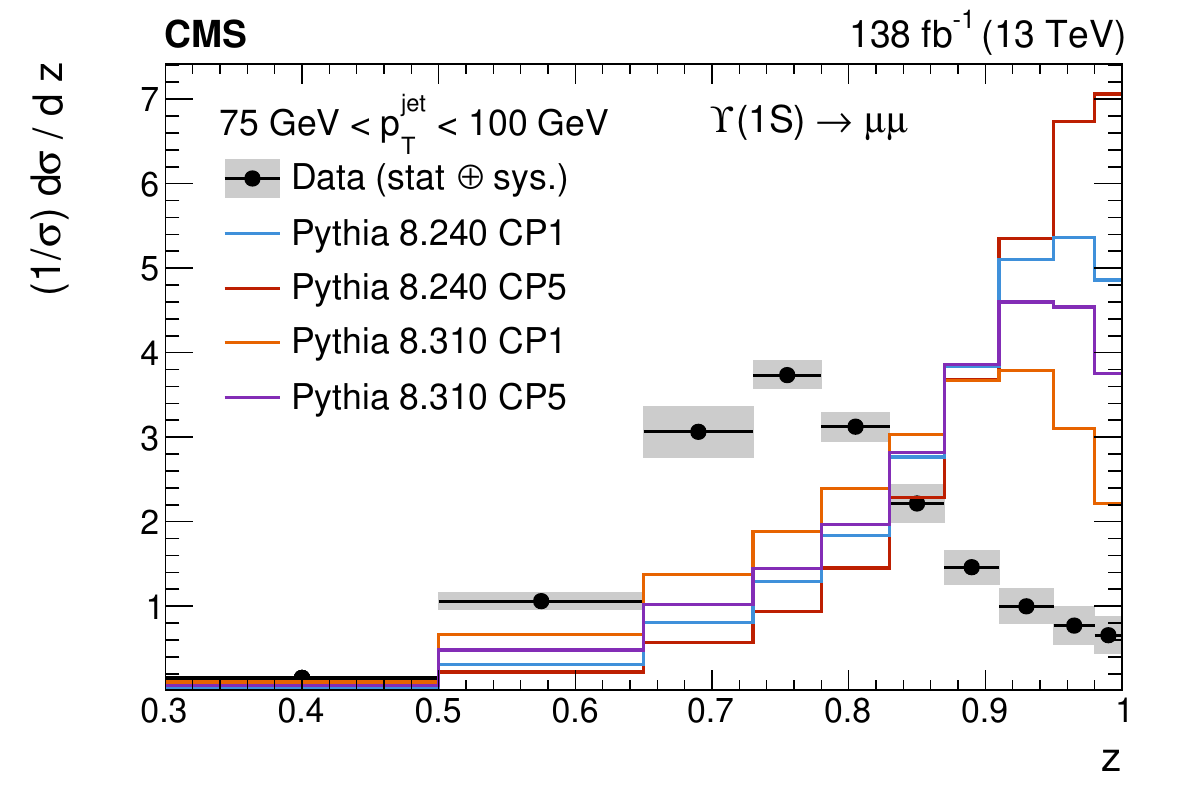}%
\hfill%
\includegraphics[width=0.48\textwidth]{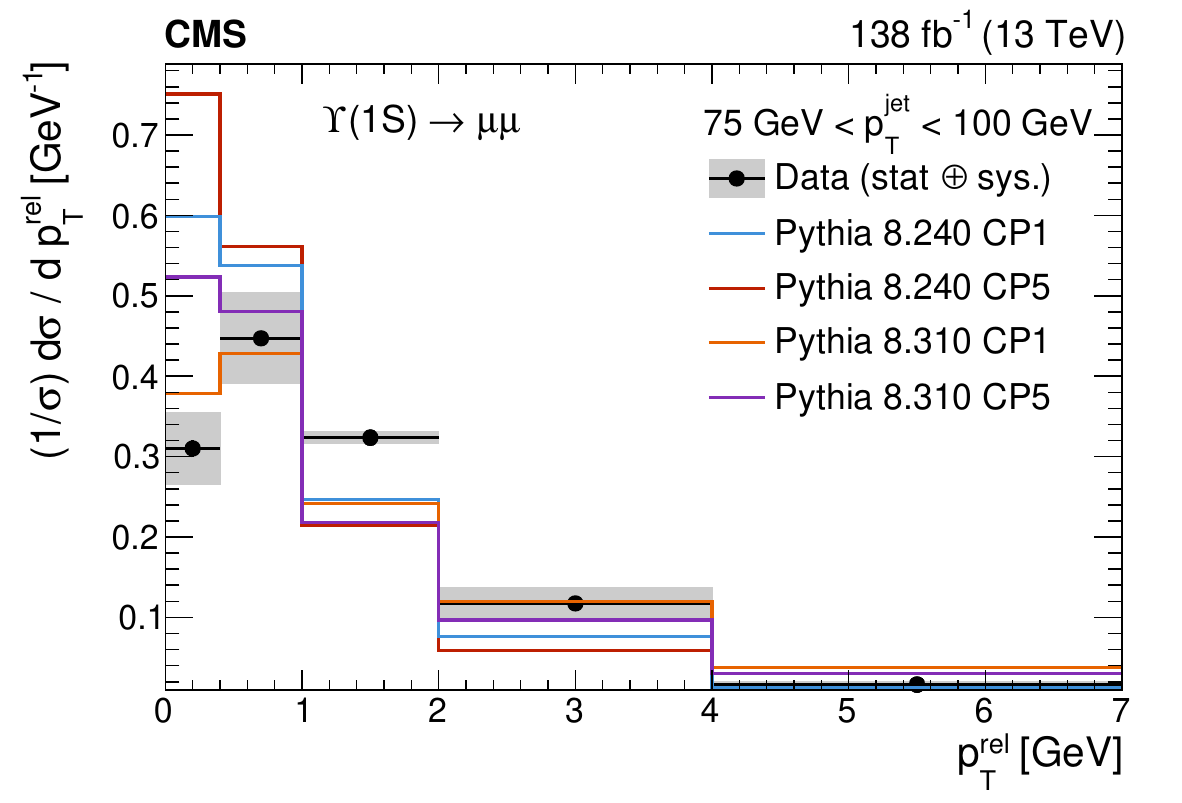} \\
\includegraphics[width=0.48\textwidth]{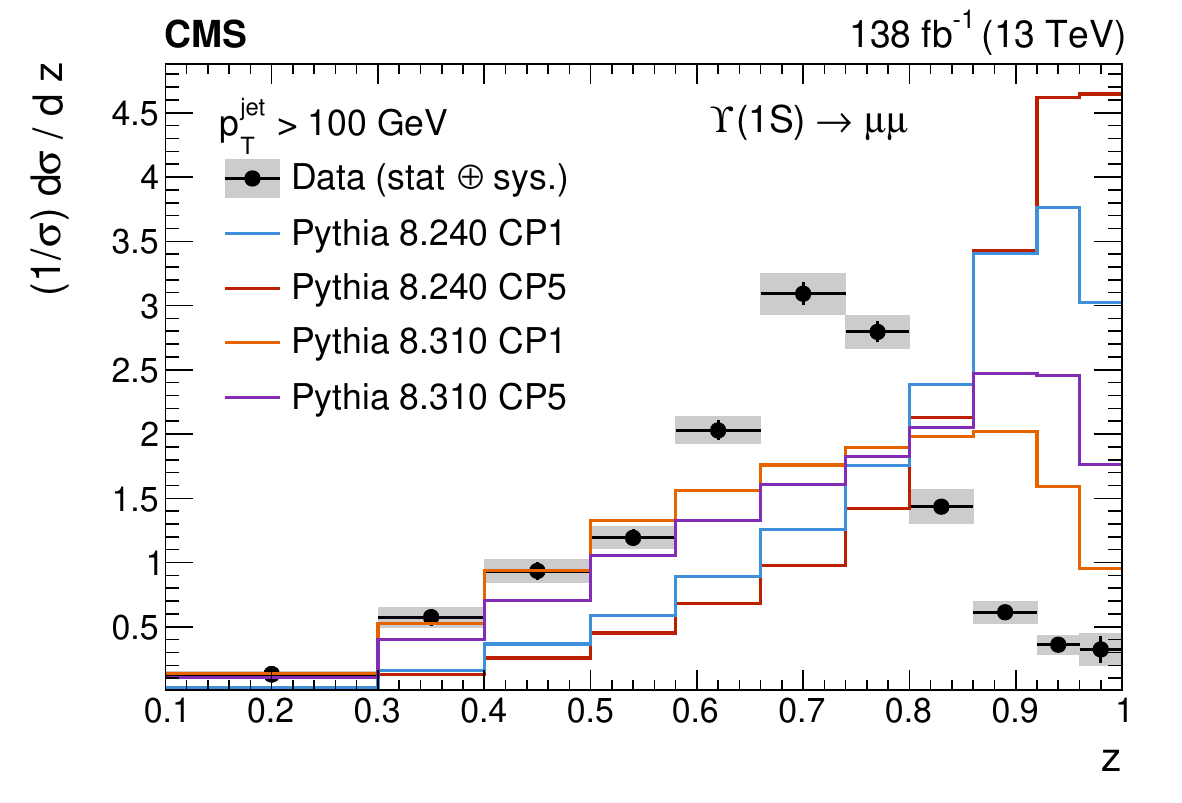}%
\hfill%
\includegraphics[width=0.48\textwidth]{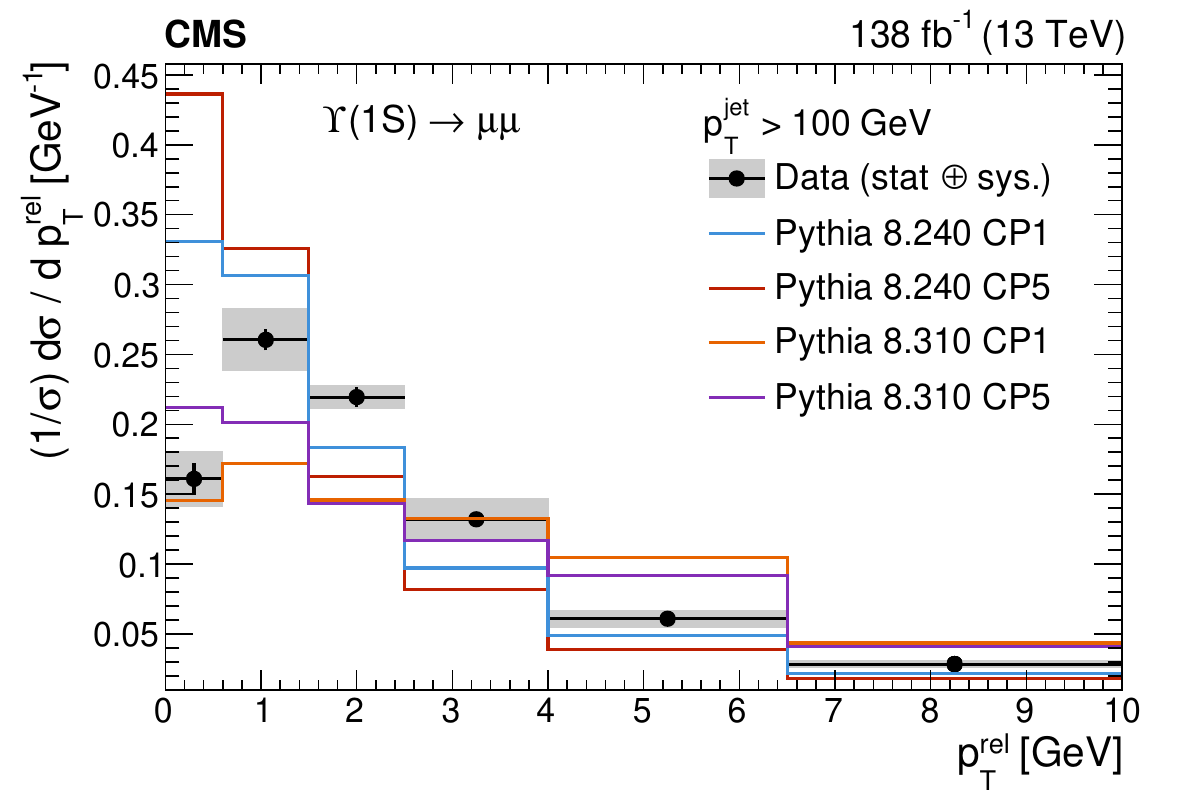}
\caption{Particle-level results for the longitudinal (left) and transverse profiles (right) of jets containing \PGUP{1S} mesons in three different \pt regions. The data (black points) are compared to the expectations of \PYTHIA8.240 with the CP1 (blue) and CP5 tunes (red), as well as the \PYTHIA8.310 predictions with the CP1 (orange) and CP5 tunes (violet). The vertical error bars represent the statistical uncertainties on the data points, while the grey bands represent the total experimental uncertainties.}
\label{fig:result1S}
\end{figure}

\begin{figure}[!p]
\includegraphics[width=0.48\textwidth]{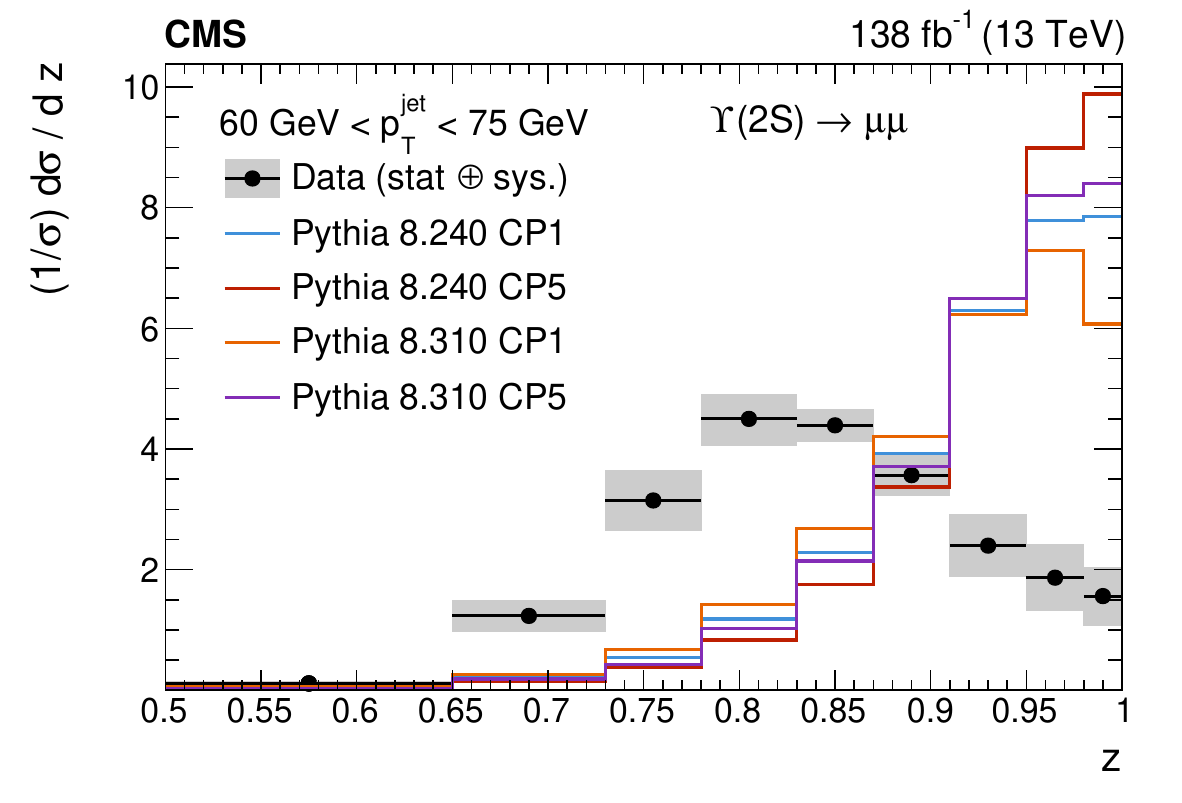}%
\hfill%
\includegraphics[width=0.48\textwidth]{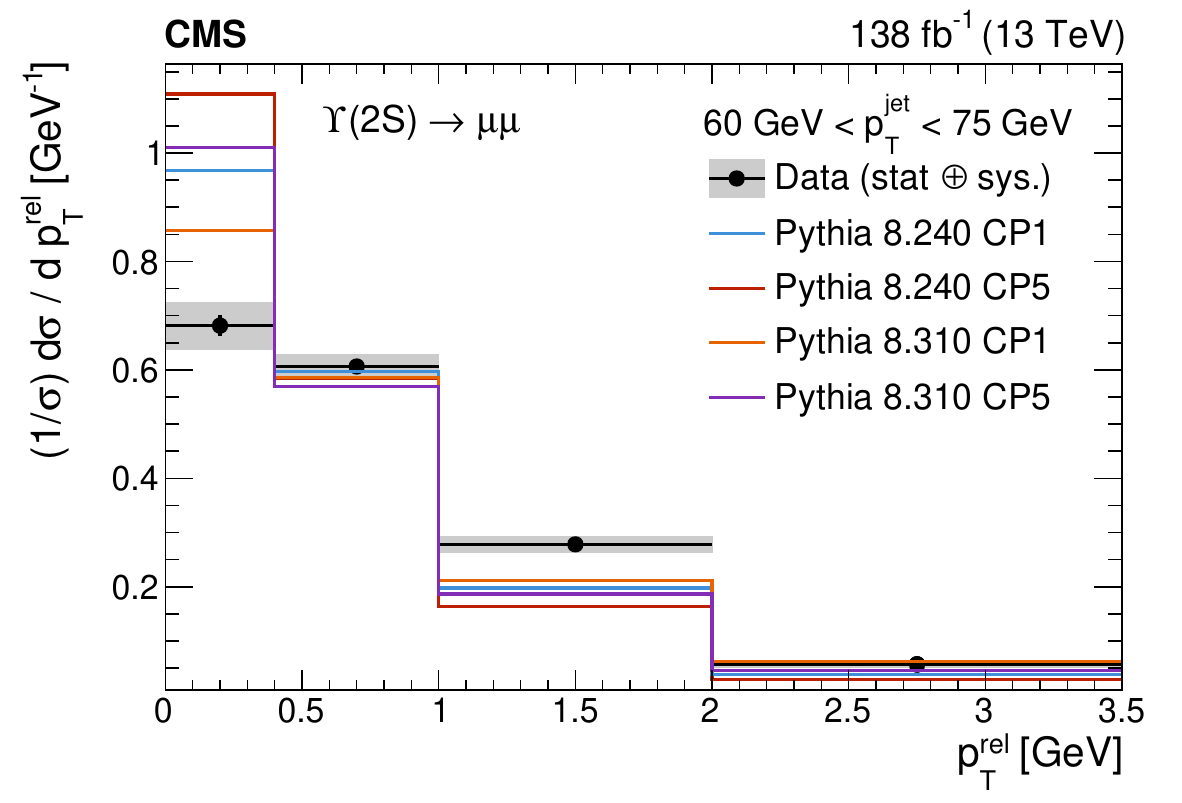} \\
\includegraphics[width=0.48\textwidth]{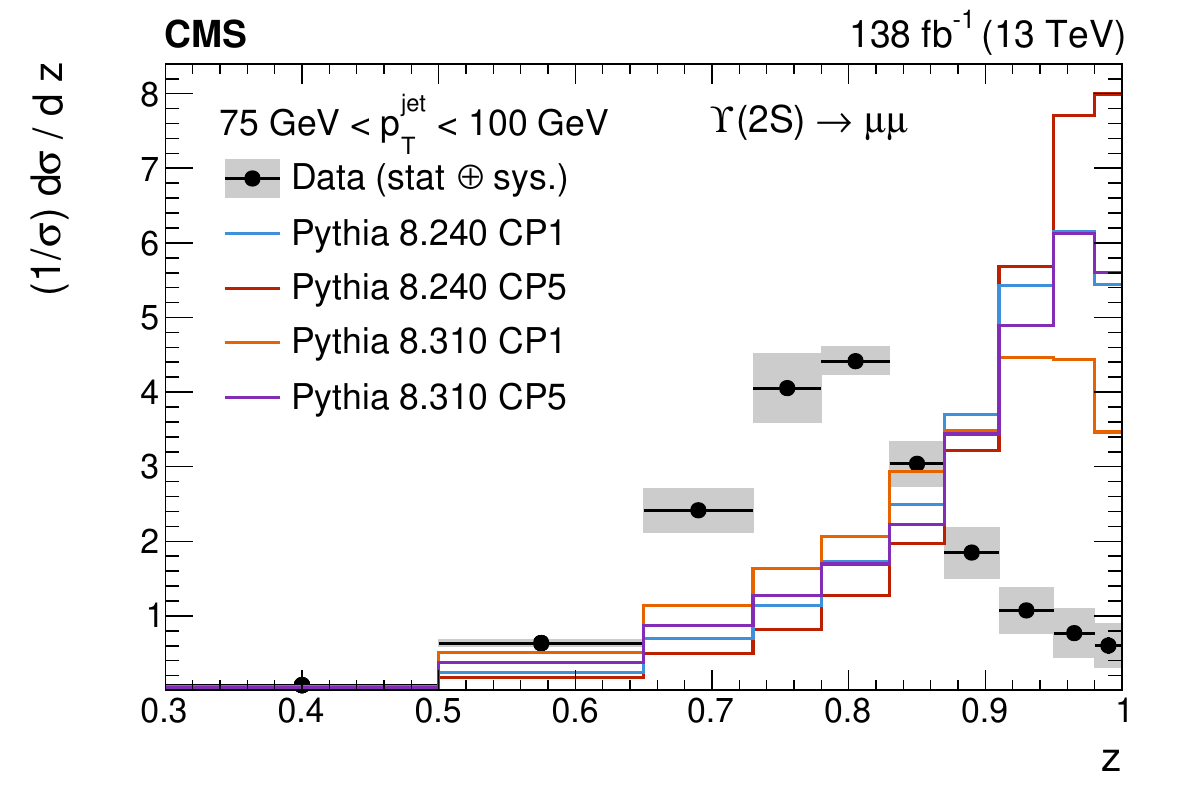}%
\hfill%
\includegraphics[width=0.48\textwidth]{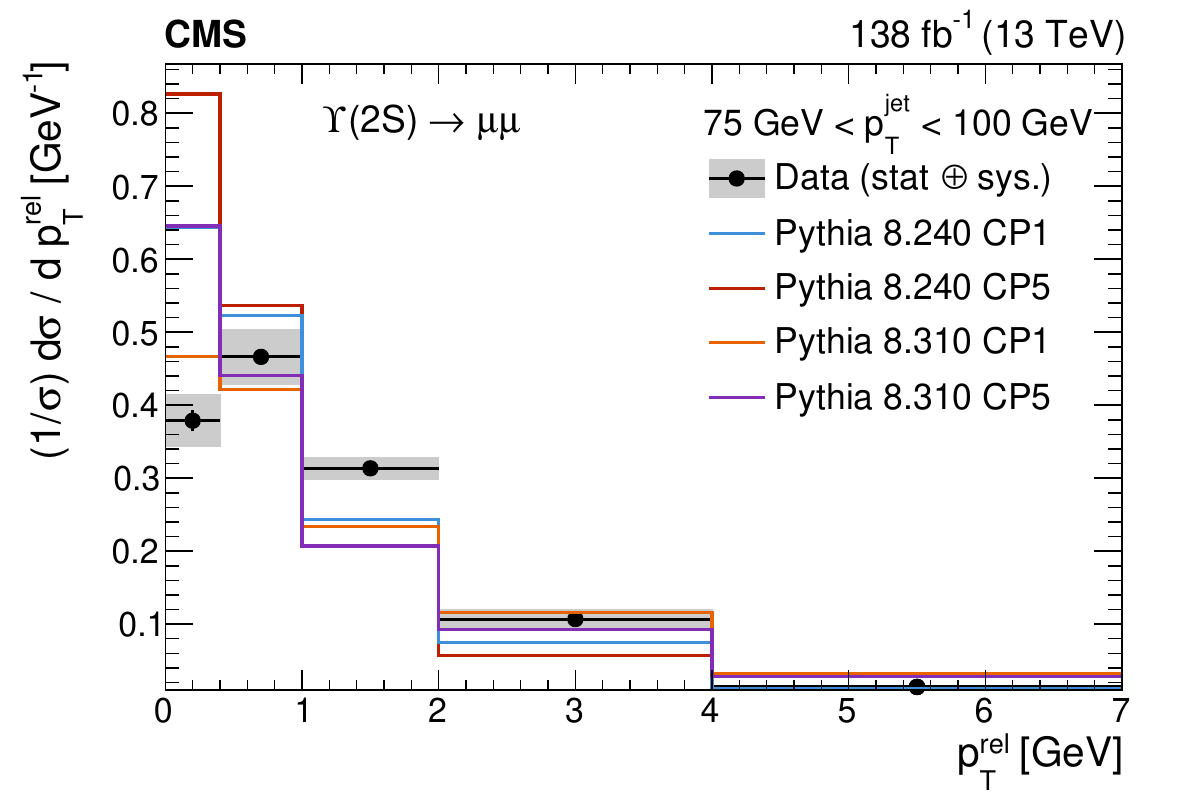} \\
\includegraphics[width=0.48\textwidth]{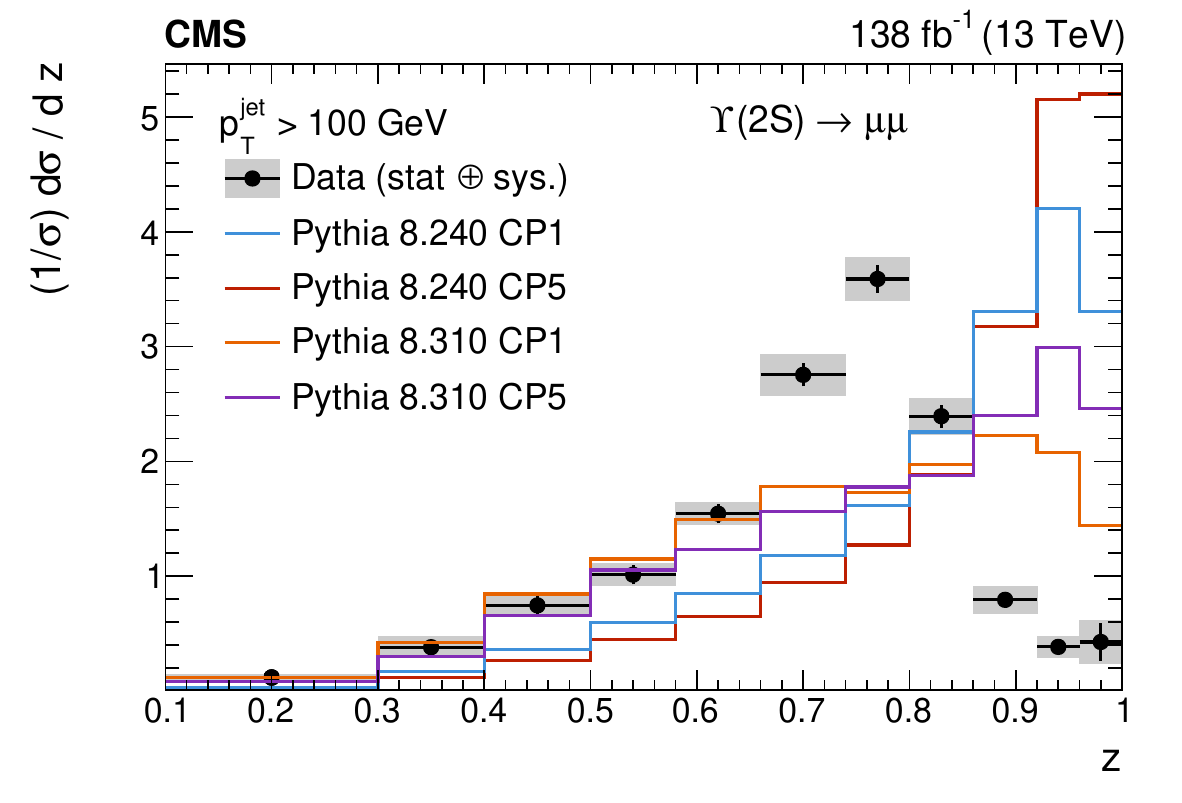}%
\hfill%
\includegraphics[width=0.48\textwidth]{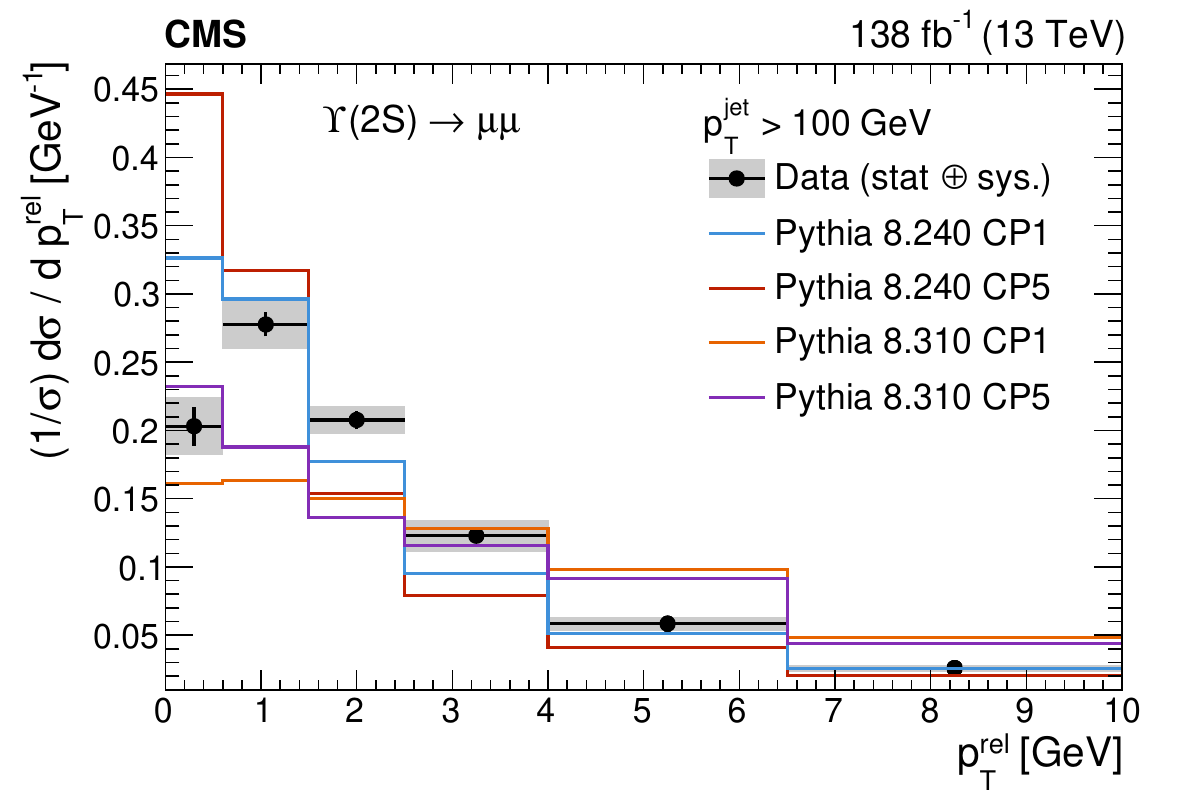}
\caption{Particle-level results for the longitudinal (left) and transverse profiles (right) of jets containing \PGUP{2S} mesons in three different \pt regions. The data (black points) are compared to the expectations of \PYTHIA8.240 with the \textsc{CP1} (blue) and \textsc{CP5} tunes (red), as well as the \PYTHIA8.310 predictions with the \textsc{CP1} (orange) and \textsc{CP5} tunes (violet). The vertical error bars represent the statistical uncertainties on the data points, while the grey bands represent the total experimental uncertainties.}
\label{fig:result2S}
\end{figure}

\begin{figure}[!t]
\includegraphics[width=0.48\textwidth]{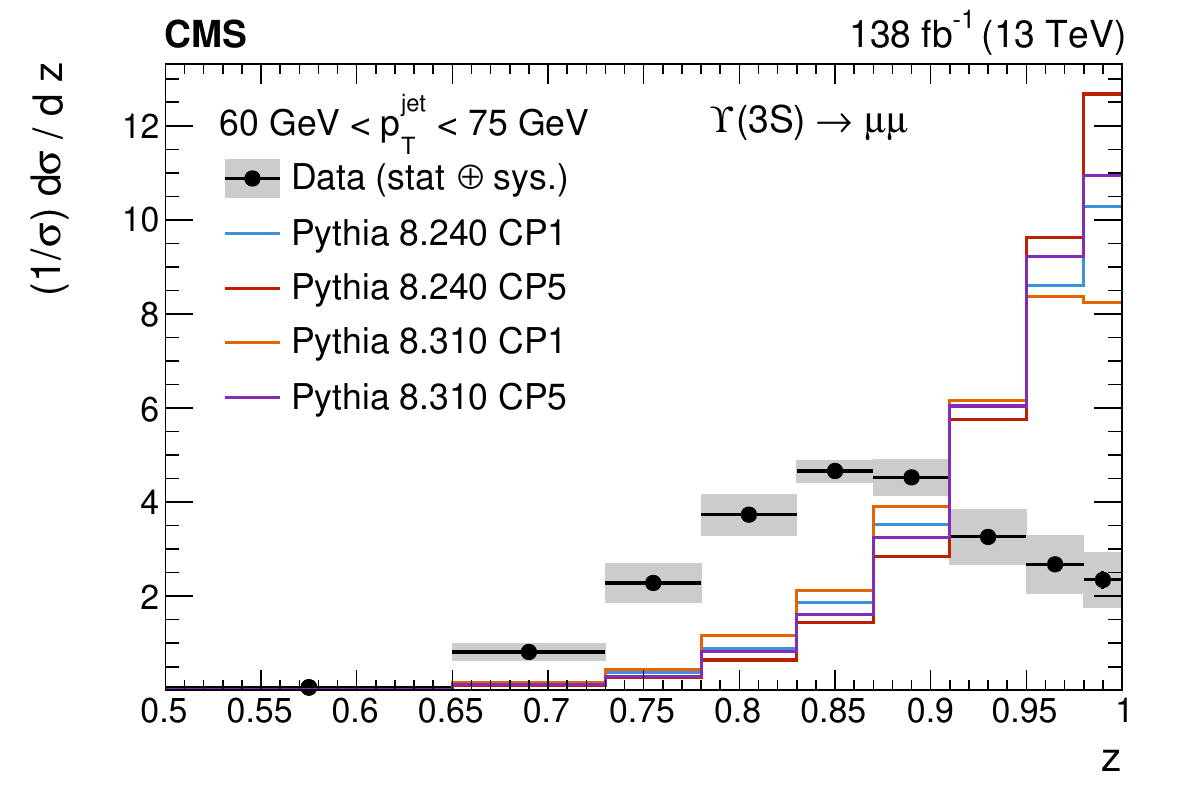}%
\hfill%
\includegraphics[width=0.48\textwidth]{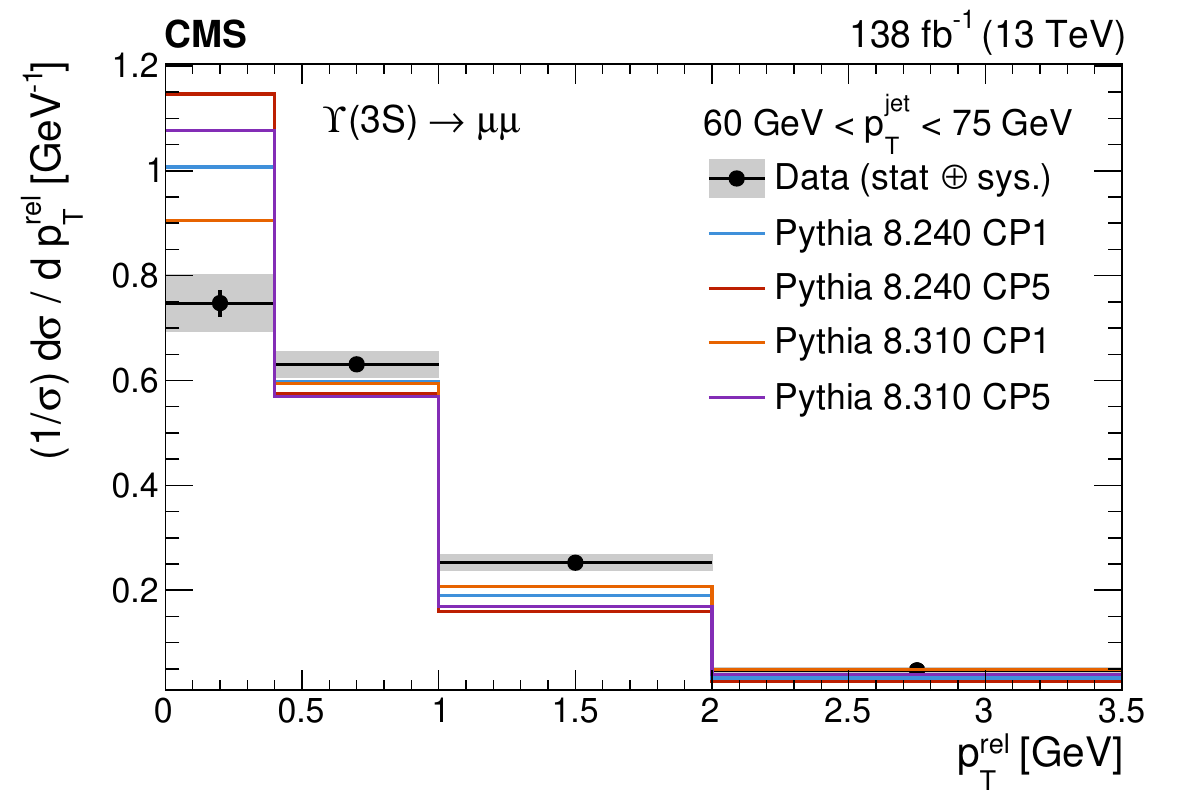} \\
\includegraphics[width=0.48\textwidth]{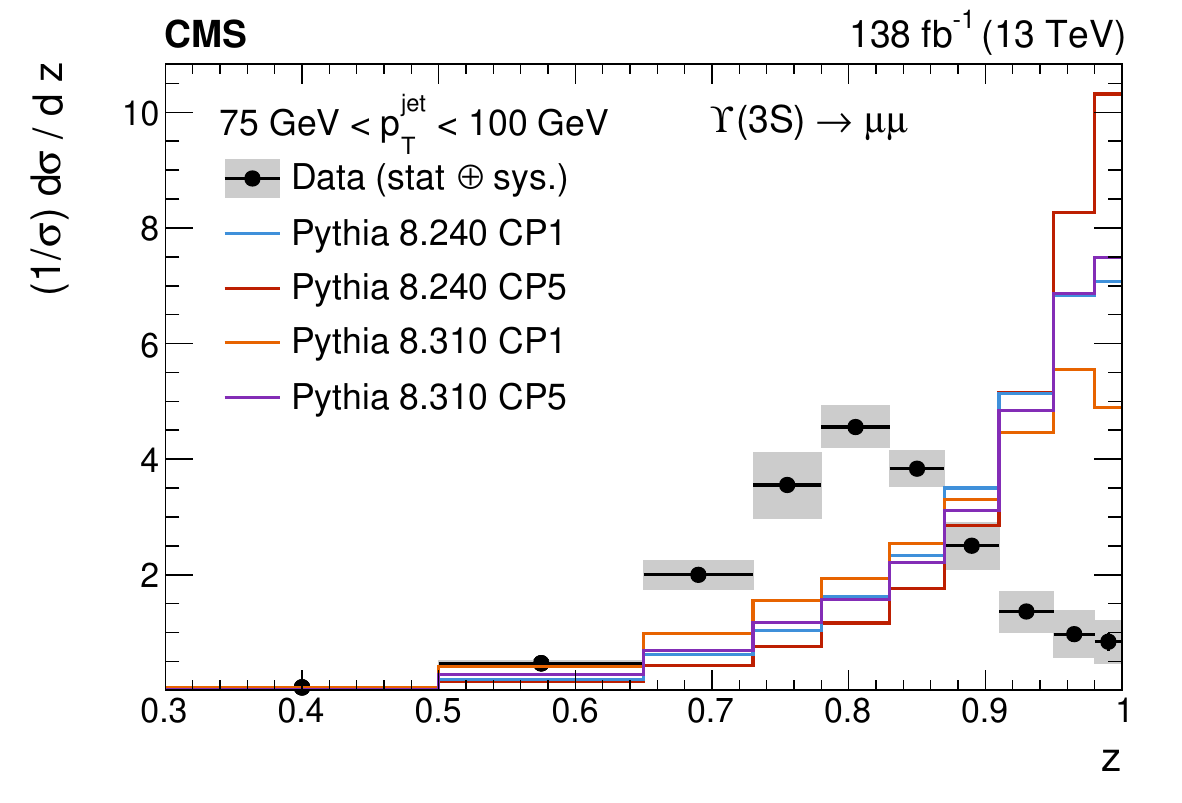}%
\hfill%
\includegraphics[width=0.48\textwidth]{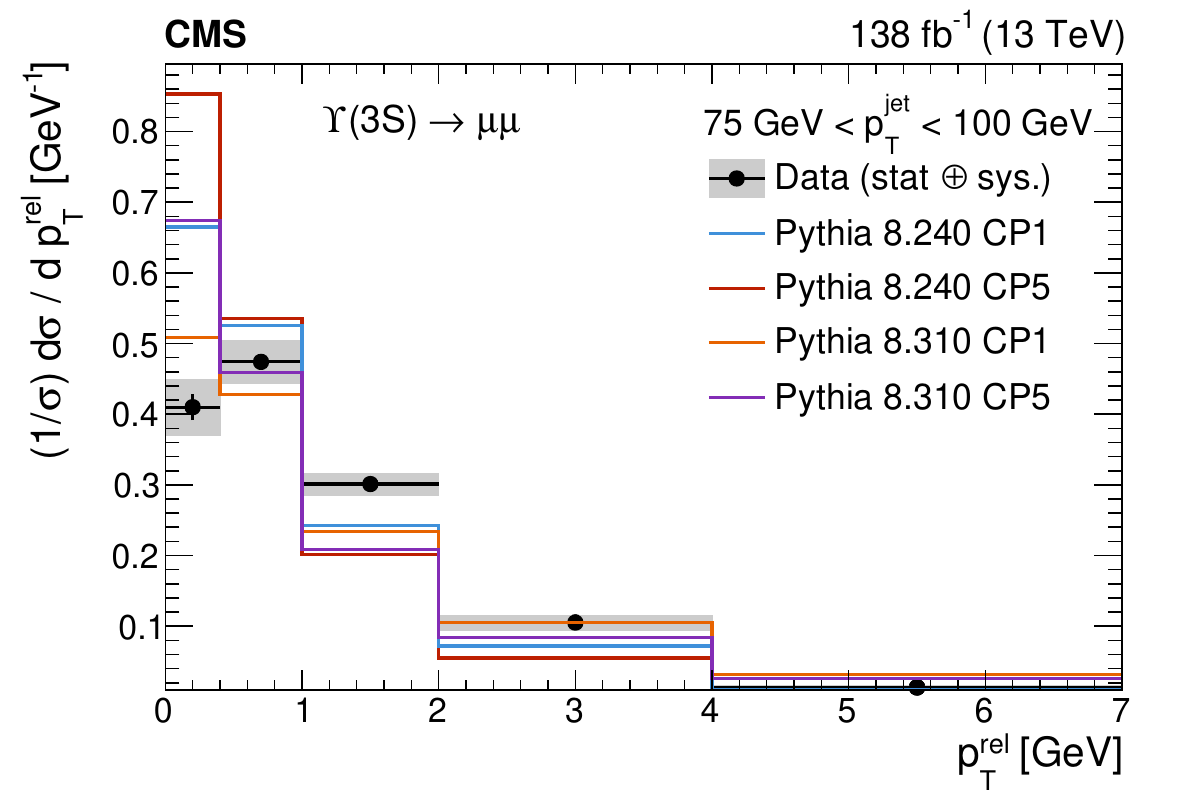} \\
\includegraphics[width=0.48\textwidth]{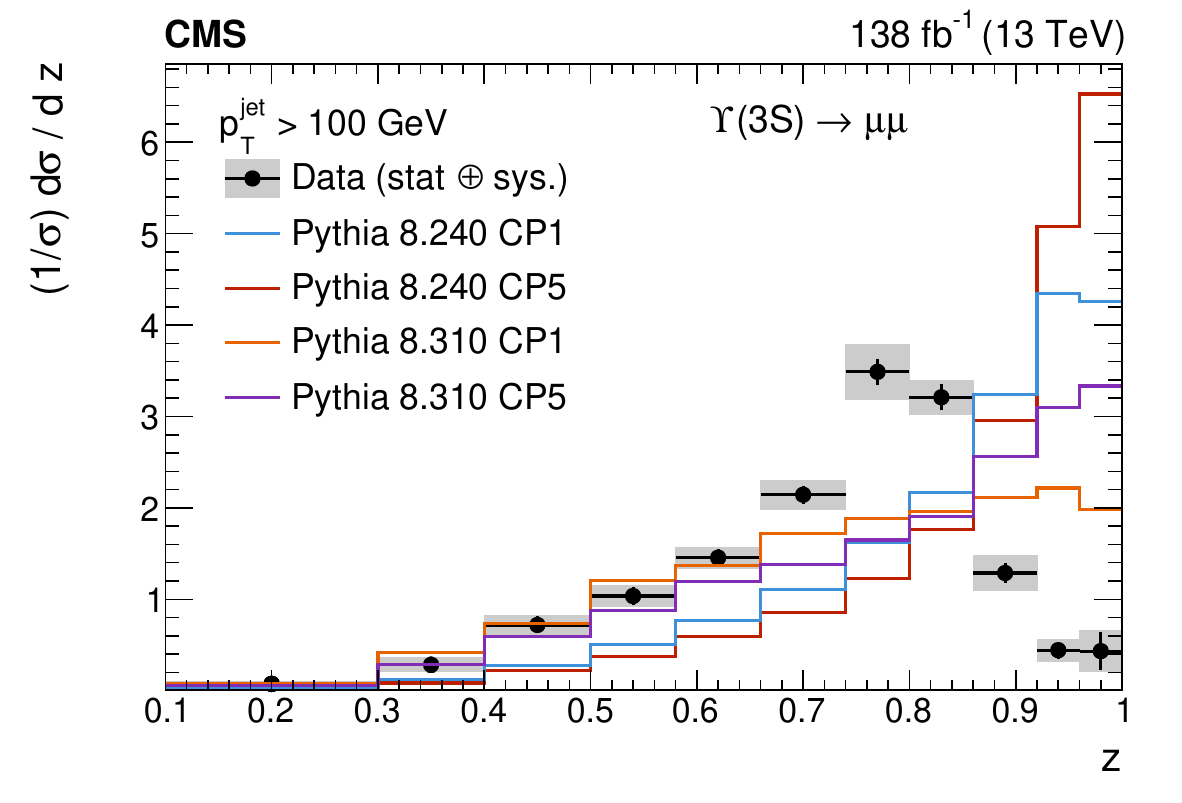}%
\hfill%
\includegraphics[width=0.48\textwidth]{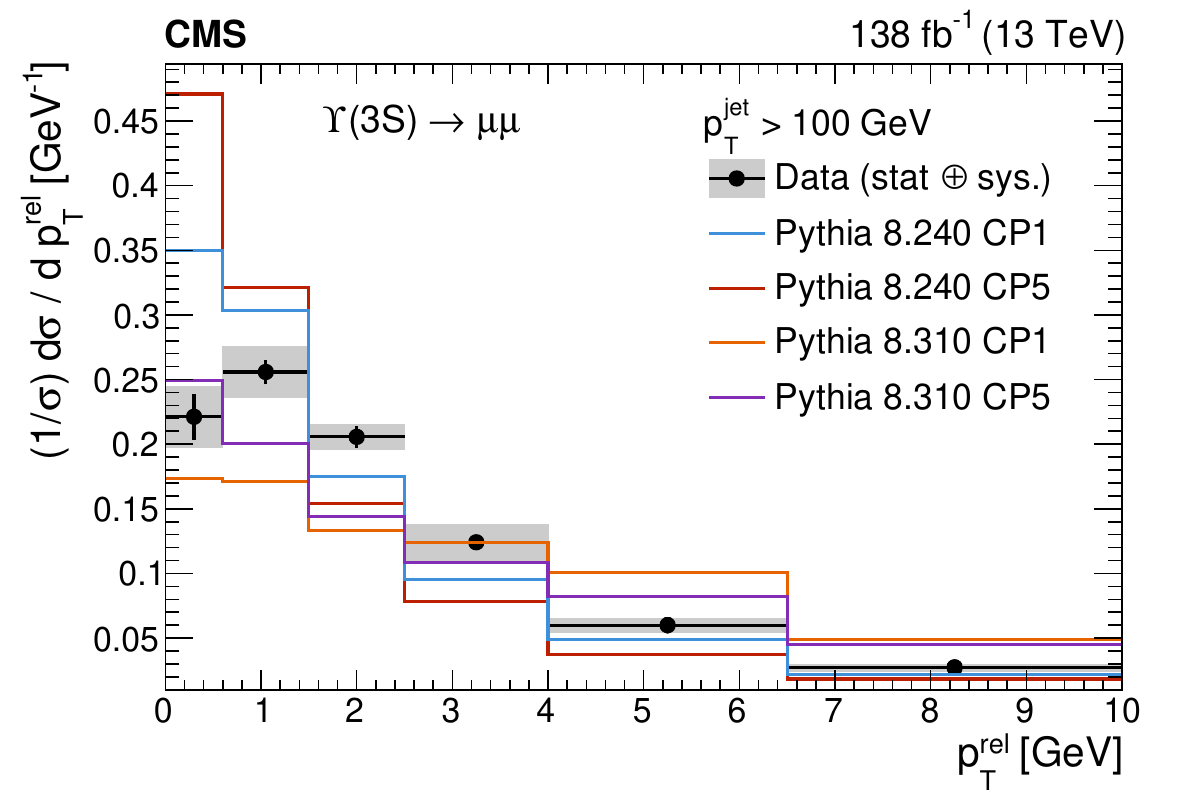}
\caption{Particle-level results for the longitudinal (left) and transverse profiles (right) of jets containing \PGUP{3S} mesons in three different \pt regions. The data (black points) are compared to the expectations of \PYTHIA8.240 with the CP1 (blue) and CP5 tunes (red), as well as the \PYTHIA8.310 predictions with the CP1 (orange) and CP5 tunes (violet). The vertical error bars represent the statistical uncertainties on the data points, while the grey bands represent the total experimental uncertainties.}
\label{fig:result3S}
\end{figure}

Figures~\ref{fig:result1S}, \ref{fig:result2S}, and \ref{fig:result3S} show the longitudinal and transverse profiles for \PGUP{1S}, \PGUP{2S}, and \PGUP{3S}, respectively, in different ranges of the jet \pt. The comparison of the data with the different MC models shows that none of the predictions described in Section~\ref{sec:montecarlo} is able to reproduce the data. In general, the MC programs under study predict more isolated \PGU mesons than observed in data, where these hadrons carry a significantly smaller fraction of the longitudinal jet momentum. A complementary effect is observed for the transverse profile, for which MC expectations generally predict a smaller \PGU meson transverse momentum wih respect to the jet axis than what is observed in data. While the improvements implemented in \PYTHIA8.310 produce less isolated quarkonia, \ie with larger lower tails for the $z$ distributions, the prediction is still far from the data points.

Small, but clear, differences in the description are observed between the CP1 and CP5 tunes. Since CP1 uses a larger value of \alpS for the modelling of the initial and final state radiation, it tends to generate additional hadronic activity, making the \PGU mesons less isolated on average and, hence, lowering the value of $z$ and increasing the value of \ptrel.

\section{Summary}
\label{sec:summary}

A measurement of the fragmentation properties of jets containing \PGUPnS quarkonia has been presented for the first time. The measurement is performed using the dimuon decays of the \PGU mesons and the jets containing the dimuon system. The measurements include the longitudinal profile $z$, the distribution of the fraction of the jet momentum carried by the \PGU meson in the direction of the jet axis, and the transverse profile \ptrel, the distribution of the momentum of the \PGU meson in the direction perpendicular to the jet axis.

The results are compared with the predictions from \PYTHIA8.240, in which \PGU mesons are generated directly from the matrix element, as well as with the \PYTHIA8.310 predictions, using an alternative approach for the generation of quarkonia from the parton shower. Both \PYTHIA versions make use of the CP1 and CP5 tunes.

These measurements provide new and stringent tests of non-relativistic QCD (NRQCD). Compared with charmonium, bottomonium production is expected to be more reliably described within the NRQCD framework, because of the larger mass of the \PQb quark, which improves the separation between perturbative and non-perturbative scales. Moreover, the inclusion of the transverse momentum profile of the \PGU meson relative to the jet axis further extends the sensitivity of these measurements to transverse-momentum-dependent aspects of quarkonium production, providing additional constraints on the dynamics of quarkonium formation beyond collinear fragmentation approaches.

The results show important discrepancies between the data and the predictions, pointing to \PGU mesons carrying a smaller momentum fraction of the jet than what is predicted in the \PYTHIA implementations of NRQCD. These results are consistent in magnitude with previous measurements of jets containing \PJGy mesons by the LHCb and CMS Collaborations. While the \PYTHIA8.310 improvements tend to go in the correct direction, showing less isolated quarkonia, these results suggest that further theoretical progress is needed to achieve a satisfactory description.

Overall, the present measurement is expected to provide additional input to the so-called quarkonium polarization puzzle by constraining the long-distance matrix elements relevant to \PGUPnS production in NRQCD. In particular, the simultaneous sensitivity to the longitudinal and transverse structure of \PGUPnS production in jets may help to disentangle the interplay between colour-singlet and colour-octet production mechanisms in this context, as well as the contributions from higher angular momentum Fock states.

\begin{acknowledgments}
We congratulate our colleagues in the CERN accelerator departments for the excellent performance of the LHC and thank the technical and administrative staffs at CERN and at other CMS institutes for their contributions to the success of the CMS effort. In addition, we gratefully acknowledge the computing centres and personnel of the Worldwide LHC Computing Grid and other centres for delivering so effectively the computing infrastructure essential to our analyses. Finally, we acknowledge the enduring support for the construction and operation of the LHC, the CMS detector, and the supporting computing infrastructure provided by the following funding agencies: SC (Armenia), BMFWF and FWF (Austria); FNRS and FWO (Belgium); CNPq, CAPES, FAPERJ, FAPERGS, and FAPESP (Brazil); MES and BNSF (Bulgaria); CERN; CAS, MoST, and NSFC (China); MINCIENCIAS (Colombia); MSES and CSF (Croatia); RIF (Cyprus); SENESCYT (Ecuador); ERC PRG and PSG, TARISTU24-TK10 and MoER TK202 (Estonia); Academy of Finland, MEC, and HIP (Finland); CEA and CNRS/IN2P3 (France); SRNSF (Georgia); BMFTR, DFG, and HGF (Germany); GSRI (Greece); MATE and NKFIH (Hungary); DAE and DST (India); IPM (Iran); SFI (Ireland); INFN (Italy); MSIT and NRF (Republic of Korea); MES (Latvia); LMTLT (Lithuania); MOE and UM (Malaysia); BUAP, CINVESTAV, CONACYT, LNS, SEP, and UASLP-FAI (Mexico); MOS (Montenegro); MBIE (New Zealand); PAEC (Pakistan); MSHE, NSC, and NAWA (Poland); FCT (Portugal);  MESTD (Serbia); MICIU/AEI and PCTI (Spain); MOSTR (Sri Lanka); Swiss Funding Agencies (Switzerland); MST (Taipei); MHESI (Thailand); TUBITAK and TENMAK (T\"{u}rkiye); NASU (Ukraine); STFC (United Kingdom); DOE and NSF (USA).

\begin{sloppypar}
\setlength\emergencystretch{\hsize}
\hyphenation{Rachada-pisek} Individuals have received support from the Marie-Curie programme and the European Research Council and Horizon 2020 Grant, contract Nos.\ 675440, 724704, 752730, 758316, 765710, 824093, 101115353, 101002207, 101001205, and COST Action CA16108 (European Union); the Leventis Foundation; the Alfred P.\ Sloan Foundation; the Alexander von Humboldt Foundation; the Science Committee, project no. 22rl-037 (Armenia); the Fonds pour la Formation \`a la Recherche dans l'Industrie et dans l'Agriculture (FRIA) and Fonds voor Wetenschappelijk Onderzoek contract No. 1228724N (Belgium); the Beijing Municipal Science \& Technology Commission, No. Z191100007219010, the Fundamental Research Funds for the Central Universities, the Ministry of Science and Technology of China under Grant No. 2023YFA1605804, the Natural Science Foundation of China under Grant No. 12535004, and USTC Research Funds of the Double First-Class Initiative No.\ YD2030002017 (China); the Ministry of Education, Youth and Sports (MEYS) of the Czech Republic; the Shota Rustaveli National Science Foundation (Georgia); the Deutsche Forschungsgemeinschaft (DFG), among others, under Germany's Excellence Strategy -- EXC 2121 ``Quantum Universe" -- 390833306, and under project number 400140256 - GRK2497; the Hellenic Foundation for Research and Innovation (HFRI), Project Number 2288 (Greece); the Hungarian Academy of Sciences, the New National Excellence Program - \'UNKP, the NKFIH research grants K 131991, K 138136, K 143460, K 143477, K 147557, K 146913, K 146914, K 147048, TKP2021-NKTA-64, and 2025-1.1.5-NEMZ\_KI-2025-00004, and MATE KKP and KKPCs Research Excellence and Flagship Research Groups grants (Hungary); the Council of Science and Industrial Research, India; ICSC -- National Research Centre for High Performance Computing, Big Data and Quantum Computing, FAIR -- Future Artificial Intelligence Research, and CUP I53D23001070006 (Mission 4 Component 1), funded by the NextGenerationEU program, the Italian Ministry of University and Research (MUR) under Bando PRIN 2022 -- CUP I53C24002390006, PRIN PRIMULA 2022RBYK7T (Italy); the Latvian Council of Science; the Ministry of Science and Higher Education, project no. 2022/WK/14, and the National Science Centre, contracts Opus 2021/41/B/ST2/01369, 2021/43/B/ST2/01552, 2023/49/B/ST2/03273, and the NAWA contract BPN/PPO/2021/1/00011 (Poland); the Funda\c{c}\~ao para a Ci\^encia e a Tecnologia (Portugal); the National Priorities Research Program by Qatar National Research Fund; MICIU/AEI/10.13039/501100011033, ERDF/EU, ``European Union NextGenerationEU/PRTR", projects PID2022-142604OB-C21, PID2022-139519OB-C21, PID2023-147706NB-I00, PID2023-148896NB-I00, PID2023-146983NB-I00, PID2023-147115NB-I00, PID2023-148418NB-C41, PID2023-148418NB-C42, PID2023-148418NB-C43, PID2023-148418NB-C44, PID2024-158190NB-C22, RYC2021-033305-I, RYC2024-048719-I, CNS2023-144781, CNS2024-154769 and Plan de Ciencia, Tecnolog{\'i}a e Innovaci{\'o}n de Asturias, Spain; the Chulalongkorn Academic into Its 2nd Century Project Advancement Project, the National Science, Research and Innovation Fund program IND\_FF\_68\_369\_2300\_097, and the Program Management Unit for Human Resources \& Institutional Development, Research and Innovation, grant B39G680009 (Thailand); the Eric \& Wendy Schmidt Fund for Strategic Innovation through the CERN Next Generation Triggers project under grant agreement number SIF-2023-004; the Kavli Foundation; the Nvidia Corporation; the SuperMicro Corporation; the Welch Foundation, contract C-1845; and the Weston Havens Foundation (USA).
\end{sloppypar}
\end{acknowledgments}\section*{Data availability} Release and preservation of data used by the CMS Collaboration as the basis for publications is guided by the  \href{https://doi.org/10.7483/OPENDATA.CMS.1BNU.8V1W}{CMS data preservation, re-use and open access policy}.

\bibliography{auto_generated}
\cleardoublepage \appendix\section{The CMS Collaboration \label{app:collab}}\begin{sloppypar}\hyphenpenalty=5000\widowpenalty=500\clubpenalty=5000\cmsinstitute{Yerevan Physics Institute, Yerevan, Armenia}
{\tolerance=6000
A.~Gevorgyan\cmsorcid{0000-0003-2751-9489}, A.~Hayrapetyan, V.~Makarenko\cmsorcid{0000-0002-8406-8605}, A.~Tumasyan\cmsAuthorMark{1}\cmsorcid{0009-0000-0684-6742}
\par}
\cmsinstitute{Marietta Blau Institute for Particle Physics, Vienna, Austria}
{\tolerance=6000
W.~Adam\cmsorcid{0000-0001-9099-4341}, L.~Benato\cmsorcid{0000-0001-5135-7489}, T.~Bergauer\cmsorcid{0000-0002-5786-0293}, M.~Dragicevic\cmsorcid{0000-0003-1967-6783}, S.~Gundacker\cmsorcid{0000-0003-2087-3266}, A.K.~Guven\cmsorcid{0009-0004-5670-5138}, P.S.~Hussain\cmsorcid{0000-0002-4825-5278}, M.~Jeitler\cmsAuthorMark{2}\cmsorcid{0000-0002-5141-9560}, N.~Krammer\cmsorcid{0000-0002-0548-0985}, A.~Li\cmsorcid{0000-0002-4547-116X}, D.~Liko\cmsorcid{0000-0002-3380-473X}, M.~Matthewman, A.~Pfeiffer\cmsorcid{0000-0001-5328-448X}, J.~Schieck\cmsAuthorMark{2}\cmsorcid{0000-0002-1058-8093}, R.~Sch\"{o}fbeck\cmsAuthorMark{2}\cmsorcid{0000-0002-2332-8784}, M.~Shooshtari\cmsorcid{0009-0004-8882-4887}, M.~Sonawane\cmsorcid{0000-0003-0510-7010}, N.~Van~Den~Bossche\cmsorcid{0000-0003-2973-4991}, W.~Waltenberger\cmsorcid{0000-0002-6215-7228}, C.E.~Wulz\cmsAuthorMark{2}\cmsorcid{0000-0001-9226-5812}
% ...
\par}
$^{1}$Also at Yerevan State University, Yerevan, Armenia\\
$^{2}$Also at Technische Universit\"{a}t Wien, Vienna, Austria\\
% ...
\end{sloppypar}
\end{document}